\documentclass[aps,pra,floats,floatfix,showpacs,showkeys,twocolumn]{revtex4}
\usepackage{amssymb}

\usepackage{epsfig}
\usepackage{graphicx}
\usepackage{amsmath}

\def\be{\begin{equation}}
\def\ee{\end{equation}}
\def\bea{\begin{eqnarray}}
\def\eea{\end{eqnarray}}
\def\bml{\begin{mathletters}}
\def\eml{\end{mathletters}}

\def\ignore#1{ }

\begin{document}

\title{Pulse-driven near-resonant quantum adiabatic dynamics: lifting of
quasi-degeneracy}
\author{L. P. Yatsenko}
\email{yatsenko@iop.kiev.ua}
\altaffiliation[Permanent address: ]{Institute~of~Physics,~Ukrainian~Academy~of~Sciences,
~prospect~Nauki~46,~Kiev-22,~252650,~Ukraine}
\author{S.~Gu\'erin}
\email{sguerin@u-bourgogne.fr}
\author{H. R.~Jauslin}
\affiliation{Laboratoire de Physique, UMR CNRS 5027, Universit\'e de Bourgogne, BP 47870,
21078 Dijon, France}
\date{\today }

\begin{abstract}
We study the quantum dynamics of a two-level system driven by a pulse that
starts near-resonant for small amplitudes, yielding nonadiabatic evolution,
and induces an adiabatic evolution for larger amplitudes. This problem is
analyzed in terms of lifting of degeneracy for rising amplitudes. It is
solved exactly for the case of linear and exponential rising. Approximate
solutions are given in the case of power law rising. This allows us to
determine approximative formulas for the lineshape of resonant excitation by
various forms of pulses such as truncated trig-pulses. We also analyze and
explain the various superpositions of states that can be obtained by the
Half Stark Chirped Rapid Adiabatic Passage (Half-SCRAP) process.
\end{abstract}

 \maketitle

\section{Introduction}

Coherent superposition of states is a key concept of comtemporary quantum
physics, for example in quantum communication and computing through
entangled states (see e.g. \cite{entang}). It is well known that various
superpositions of states can be created in a two-level atom (states $\left|
-\right\rangle $ and $\left| +\right\rangle $) travelling through a laser
(or equivalently for a cold atom driven by a pulsed laser) or through a
cavity. If the population initially resides in the ground state $\left|
-\right\rangle $, a laser-induced one-photon resonant process leads to a
superposition at the end of the interaction [in the rotating wave
approximation (RWA)] of the form%
\begin{equation}
\left| \psi \right\rangle =\cos \int_{t_{i}}^{t_{f}}\frac{\Omega (t)}{2}dt\
\left| -\right\rangle -ie^{-i\omega t}\sin \int_{t_{i}}^{t_{f}}\frac{\Omega
(t)}{2}dt\ \left| +\right\rangle ,  \label{superp}
\end{equation}%
where $\Omega (t)$ is the Rabi frequency (assumed real positive)
proportional to the pulse envelope and the coupling, that is integrated
between $t_{i}$ and $t_{f},$ the initial and final times of interaction, and
$\omega $ is the optical frequency. In a one-photon resonant cavity,
prepared in a Fock state $\left| n+1\right\rangle $, $n\geq 0$, the
counterpart of Eq. (\ref{superp}) reads%
\begin{align}
\left| \psi \right\rangle & =\cos \left( \sqrt{n+1}\int_{t_{i}}^{t_{f}}\frac{%
\Omega (t)}{2}dt\right) \left| -,n+1\right\rangle  \notag \\
& -i\sin \left( \sqrt{n+1}\int_{t_{i}}^{t_{f}}\frac{\Omega (t)}{2}dt\right)
\left| +,n\right\rangle ,  \label{superp2}
\end{align}%
where $\left| i,k\right\rangle $ represents the atomic bare state $\left|
i\right\rangle $ dressed by $k$ photons of the cavity. For instance, a half
superposition is achieved when the Rabi frequency area is $\pi /2$. From the
practical point of view, this process requires the precise control of the
effectively interacting pulse shape, i.e. in the case of a travelling atom,
this requires a well-controlled and homogenous velocity and the control of
the characteristics of the intersection between laser and atomic beam. As a
consequence, this process is said to be non-robust with respect to the pulse
amplitude. Creation of coherent superpositions of states by adiabatic
passage is of great interest since this process requires a sufficiently
large pulse area, but not precisely defined: hence it is robust with respect
to fluctuations of the interaction and also to partial knowledge of atomic
and field parameters.

The analysis of adiabatic evolution and its leading corrections is
well-understood in the case of exact (one-photon) resonance and in the far
from resonance case. The behavior in theses two exterme cases is
qualitatively different. As a consequence, in the intermediate regime that
we will characterize as quasi-resonant, there is a competition of different
effects that to our knowledge has not been studied quantitatively. The goal
of this article is to present a detailled analysis of this intermediate
regime, whose understanding is important for practical applications. We
obtain the results combining insights from exactly solvable models with
approximations obtained by perturbative techniques in different regimes. We
test the quantitative validity of our results by comparison with direct
numerical simulations.

We can identify the different known regimes of the dynamics in a two-level
system considering a detuning $\Delta $. We denote by $T$ the characteristic
time of the rising (and falling) of the pulse.
The resonant case is defined as%
\begin{equation}
T\Delta \ll 1.  \label{resonant}
\end{equation}%
We can reintepret the formulas (\ref{superp}) and (\ref{superp2}) and extend
them to the case of multiphoton resonant processes with adiabatic pulses as
follows \cite{Holthaus,Guerin_PRA97,review}. In the case of an exact $n-$%
photon resonance, the two relevant dressed states $\left| -,n\right\rangle $
and $\left| +,0\right\rangle $ can be considered, before the rising of the
pulse, as exactly degenerate with respect to the dynamics. The pulse rising
induces a\textit{\ lifting of the degeneracy}, which leads to a \textit{%
splitting} of the dynamics along the two eigenstate branches. It is
important to note that this splitting is \textit{instantaneous} at the
beginning of the pulse only in the case of one-photon resonance $n=1$ and in
the two-photon case $n=2$ \cite{review}. Higher multiphoton processes
involve Stark shifts which modify the splitting. The $n=1$ case induces an
equal splitting along the two eigenstate branches. These two branches are
next \textit{followed adiabatically} by the dynamics if the pulse envelopes
are slow enough. One can define dynamical phases which are the areas between
the associated dressed eigenenergies. When later the pulse falls, the
dynamics faces the symmetrically inverse problem of the \textit{creation of
degeneracy} with a\textit{\ recombination} (instantaneous for $n=1$ and $n=2$%
) leading to the interference of the two branches at the very end of the
process. The resulting final transfer depends on (i) the way in which the
splitting (and recombination) occurs (in amplitude and phase), and (ii) the
difference of the dynamical phases. In the case of a complete transfer, the
process has been named generalized or multiphoton $\pi $-pulse. This process
is not robust because the splitting (and recombination) and the difference
of the dynamical phases depend on the parameters. Moreover, numerics shows
that it is much more sensitive to the detuning from the multiphoton
resonance than to the pulse shape \cite{Just,Korolkov}.

In the opposite regime far from the resonance defined by the condition
\begin{equation}
T\Delta \gg 1,  \label{adiabcond}
\end{equation}%
the dynamics is at all time adiabatic in the sense that it follows the
single dressed eigenstate whose eigenvalue is continuously connected to the
one associated to the initial bare state. The nonadiabatic corrections
(exponentially small for smooth pulses), that produce losses into the other
eigenstate have been extensively studied, e.g. in \cite{Berman}.

If the detuning is additionally time dependent (induced by a direct
frequency chirping or by an additional off-resonant pulse which Stark shifts
the states) and if the condition (\ref{adiabcond}) is satisfied during the
rising and the falling of the pulse, it has been shown \cite%
{Guerin_PR00,Yatsenko_PR01,review} that the topology of the dressed
eigenenergy surfaces, as functions of the effective time-dependent external
field parameters, allows to determine the various possible population
transfers. The main ingredient is a \textit{global adiabatic passage} along
\textit{one} eigenstate combined with \textit{local crossings} of resonances
which appear as conical intersections and that can be precisely determined
from the eigenenergy surfaces. This adiabatic passage results at the end of
the process either in a complete population transfer to the excited state or
in a complete return to the ground state. This process in the case of an
additional Stark laser, leading to a complete population transfer, has been
named Stark Chirped Rapid Adiabatic Passage (SCRAP) \cite%
{Yatsenko_PR99,Rickes}.

In this paper, we study the intermediate quasi-resonant regimes, defined by
the condition
\begin{equation}
T\Delta \sim 1.
\end{equation}%
They lead to a \textit{lifting (and creation) of a quasi-degeneracy}. We
construct formulas characterizing the dynamics at asymptotic times beyond
the lifting of quasi-degeneracy, assuming adiabatic evolution along the two
branches.

Resonance between two quantum states and the resulting transitions are
mainly understood through the asymptotic limit of the Landau-Zener avoided
crossing model \cite{Landau,Zener,Dykhne,Davis}: A complete transition can
be achieved by adiabatic passage beyond the avoided crossing, along the
state continuously connected to the initial one. We will show that in the
case of pulses with amplitude growing linearly in time, the problem of the
lifting of quasi-degeneracy can be interpreted as a half Landau-Zener
process \cite{Vitanov}.

We also solve the problem of the lifting of quasi-degeneracy beyond the half
Landau-Zener model, for pulses rising as power of time and as smooth
exponential ramps. 

In the next section, we describe the model with the different couplings. In
Section III, we define the adiabatic states in the model and the conditions
for adiabatic evolution along one of the adiabatic states. We show the
dynamics when the adiabatic conditions are not satisfied at early times.
Section IV and V are devoted to the calculation of the dynamics respectively
for linear and exponential rising coupling. In section VI and VII, we
analyze the dynamics with perturbation theory in the limits of respectively
large and small detuning. On the basis of the results of Sections IV, VI and
VII, we give an approximative formula in Section VIII for a power law rising
of the coupling. In Sections IX and X, we apply the results to obtain the
lineshape of pulsed resonant excitation and to the analysis of the
Half-SCRAP process. In Section XI, we present some conclusions and open
related problems.

\section{The model}

We study a two-level system (states $\left| -\right\rangle $ and $\left|
+\right\rangle $) driven by a near-resonant pulsed laser whose state
evolution $\phi (\tau )$ is given by the Schr\"{o}dinger equation
\begin{equation}
i\hbar \frac{\partial \phi }{\partial \tau }(\tau )=T_{0}\mathsf{H}(\tau
)\phi (\tau ),\quad \phi (\tau )=\left[
\begin{array}{c}
B_{-}(\tau ) \\
B_{+}(\tau )%
\end{array}%
\right] \in \mathbb{C}^{2},  \label{se}
\end{equation}%
with $\left| B_{-}(\tau )\right| ^{2}+\left| B_{+}(\tau )\right| ^{2}=1$,
the scaled time $\tau =t/T_{0}$ and the Hamiltonian in the quasi-resonant
approximation \cite{Allen,Shore}
\begin{equation}
\mathsf{H}(\tau )=\frac{\hbar }{2}\left[
\begin{array}{cc}
-\Delta (\tau ) & \Omega (\tau ) \\
\Omega (\tau ) & \Delta (\tau )%
\end{array}%
\right] ,  \label{Ham1}
\end{equation}%
where we have considered the basis $\left\{ \left| -\right\rangle ,\left|
+\right\rangle \right\} $.

We consider the Hamiltonian (\ref{Ham1}) with a constant detuning
\begin{equation}
\Delta (\tau )=\Delta _{0}>0
\end{equation}%
and the following models of coupling between the initial $\tau _{i}$ and a
final time $\tau _{f}$:

(i) power law rising%
\begin{equation}
\Omega (\tau )=\left\{
\begin{array}{c}
\Omega _{0}\tau ^{n},\;\tau \geqslant \tau _{i} \\
0,\;\tau <\tau _{i},%
\end{array}%
\right.  \label{Rabi0}
\end{equation}%
with $\tau _{i}=0,$ $\tau _{f}\gtrsim 1$, and falling%
\begin{equation}
\Omega (\tau )=\left\{
\begin{array}{c}
\Omega _{0}\left( -\tau \right) ^{n},\;\tau \leq \tau _{f} \\
0,\;\tau >\tau _{f},%
\end{array}%
\right.  \label{Rabi0_}
\end{equation}%
with $\tau _{i}\lesssim -1$, $\tau _{f}=0$, for an integer $n\geq 1$,

(ii) smooth exponential rising%
\begin{equation}
\Omega (\tau )=\Omega _{0}e^{\tau }  \label{exp_rising}
\end{equation}%
with $\tau _{i}\rightarrow -\infty ,$ $\tau _{f}\gtrsim 0$, and falling%
\begin{equation}
\Omega (\tau )=\Omega _{0}e^{-\tau }  \label{exp_falling}
\end{equation}%
with $\tau _{i}\lesssim 0$, $\tau _{f}\rightarrow +\infty ,$ and

(iii) smooth Gaussian
\begin{equation}
\Omega (\tau )=\Omega _{0}e^{-\tau ^{2}}  \label{Gaussian}
\end{equation}%
with $\tau _{i}\rightarrow -\infty $ and/or $\tau _{f}\rightarrow +\infty $.
We assume $\Omega (\tau )\geq 0$.

We moreover consider for the pulse rising the initial conditions at time $%
\tau _{i}$: $B_{-}(\tau _{i})=1$, $B_{+}(\tau _{i})=0$.

\section{Adiabatic and nonadiabatic evolution}

\subsection{Adiabatic transformation}

The adiabatic states $\Phi _{\pm }(\tau )$ are defined as the eigenstates of
$\mathsf{H}(\tau )$, associated to the eigenvalues $\lambda _{\pm }(\tau )$:%
\begin{equation}
\mathsf{H}(\tau )\Phi _{\pm }(\tau )=\lambda _{\pm }(\tau )\Phi _{\pm }(\tau
).
\end{equation}%
Gathering the adiabatic states in the columns of the unitary matrix $\mathsf{%
R}(\tau )=\left[ \Phi _{-}(\tau ),\Phi _{+}(\tau )\right] :$%
\begin{equation}
\mathsf{R}(\tau )=\left[
\begin{array}{cc}
\cos \theta (\tau ) & \sin \theta (\tau ) \\
-\sin \theta (\tau ) & \cos \theta (\tau )%
\end{array}%
\right]
\end{equation}%
with
\begin{equation}
\tan 2\theta (\tau )=\frac{\Omega (\tau )}{\Delta (\tau )},\quad 0\leq
\theta (\tau )<\pi /2,  \label{theta}
\end{equation}%
giving%
\begin{equation}
\mathsf{R}^{\dagger }(\tau )T_{0}\mathsf{H}(\tau )\mathsf{R}(\tau )=\left[
\begin{array}{cc}
\lambda _{-} & 0 \\
0 & \lambda _{+}%
\end{array}%
\right] ,\quad \lambda _{\pm }(\tau )=\pm \frac{\hbar }{2}T_{0}\delta (\tau )
\end{equation}%
with
\begin{equation}
\delta (\tau )=\sqrt{\Delta ^{2}(\tau )+\Omega ^{2}(\tau )},  \label{quasi}
\end{equation}%
we can rewrite the Schr\"{o}dinger equation as%
\begin{equation}
i\hbar \frac{\partial }{\partial \tau }\phi _{A}(\tau )=\mathsf{H}_{A}(\tau
)\phi _{A}(\tau )  \label{Schrod}
\end{equation}%
with%
\begin{equation}
\mathsf{H}_{A}(\tau )=\frac{\hbar }{2}\left[
\begin{array}{cc}
-T_{0}\delta (\tau ) & -i\gamma (\tau ) \\
i\gamma (\tau ) & T_{0}\delta (\tau )%
\end{array}%
\right] ,
\end{equation}%
the non-adiabatic coupling
\begin{equation}
\gamma (\tau )\equiv 2\frac{d\vartheta (\tau )}{d\tau }=\frac{\dot{\Omega}%
(\tau )\Delta (\tau )-\Omega (\tau )\dot{\Delta}(\tau )}{\Delta ^{2}(\tau
)+\Omega ^{2}(\tau )},
\end{equation}%
and%
\begin{equation}
\phi _{A}(\tau )\equiv \left[
\begin{array}{c}
A_{-}(\tau ) \\
A_{+}(\tau )%
\end{array}%
\right] =\mathsf{R}^{\dagger }(\tau )\phi (\tau )=\mathsf{R}^{\dagger }(\tau
)\left[
\begin{array}{c}
B_{-}(\tau ) \\
B_{+}(\tau )%
\end{array}%
\right] .  \label{phiA}
\end{equation}%
Since $A_{\pm }(\tau )\equiv \left\langle \pm \right| \left. \phi _{A}(\tau
)\right\rangle =\left\langle \Phi _{\pm }(\tau )\right| \left. \phi (\tau
)\right\rangle $, one can interpret $\left| A_{\pm }(\tau )\right| ^{2}$ as
the population of the adiabatic states $\left| \Phi _{\pm }(\tau
)\right\rangle $. In the adiabatic limit, mathematically defined as $%
T_{0}\rightarrow \infty $, the non-adiabatic coupling $\gamma $ can be
neglected and the dynamics follows the adiabatic states.

Since $\Omega (\tau )$ and $\Delta (\tau )$ are assumed positive, at times $%
\tau _{0}$ for which the pulse goes to zero, one has%
\begin{equation}
\Omega (\tau _{0})\rightarrow 0,
\end{equation}%
$\theta (\tau _{0})\rightarrow 0$ and $\phi _{A}(\tau _{0})=\phi (\tau _{0})$%
, hence
\begin{equation}
A_{-}(\tau _{0})=B_{-}(\tau _{0}),\quad A_{+}(\tau _{0})=B_{+}(\tau _{0}).
\label{init0}
\end{equation}

\subsection{Condition of adiabatic evolution}

In this section, we consider the rising of the pulse. Since one considers in
the following a constant detuning $\Delta (\tau )=\Delta _{0}$, the dynamics
is adiabatic if
\begin{equation}
\widetilde{\gamma }(\tau )\equiv \frac{|\gamma (\tau )|}{2T_{0}\lambda (\tau
)}\ll 1.
\end{equation}%
i.e.
\begin{equation}
\frac{1}{2}\frac{\dot{\Omega}(\tau )\Delta _{0}}{\left( \Delta
_{0}^{2}+\Omega ^{2}(\tau )\right) ^{3/2}}\ll T_{0},  \label{adiab_following}
\end{equation}%
where we have defined the nonadiabatic coefficient $\widetilde{\gamma }(\tau
)$ (which appears at the first order of the time-dependent perturbation
theory). \textit{We assume here that the time $t=T_{0}$ is a characteristic
time beyond which the evolution of the system is adiabatic}, i.e. without
population transfer between adiabatic states. Our goal is to find the
population transfer during the characteristic time \textit{before}
adiabaticity.

For times $\tau \sim 1$ (i.e. $t\sim T_{0}$), Eq. (\ref{adiab_following})
for a power law coupling (\ref{Rabi0}) reads
\begin{equation}
2T_{0}\left[ \Delta _{0}^{2}+\Omega _{0}^{2}\right] ^{3/2}\gg n\Omega
_{0}\Delta _{0},
\end{equation}%
which, using $\Omega _{0}\Delta _{0}/(\Delta _{0}^{2}+\Omega _{0}^{2})\sim 1$%
, can be roughly simplified as%
\begin{equation}
2T_{0}\left[ \Delta _{0}^{2}+\Omega _{0}^{2}\right] ^{1/2}\gg n.  \label{ad2}
\end{equation}%
We assume in this paper that this condition (\ref{ad2}) is satisfied for
power law couplings. For \textit{large detunings} defined here as $%
T_{0}\Delta _{0}\gg n$, the evolution is adiabatic around these times $\tau
\sim 1$ for any $\Omega _{0}$ (and is actually adiabatic at anytime if one
additionally excludes $\Omega _{0}\gg \Delta _{0}$,\ as shown below). For
\textit{intermediate detunings} defined as $T_{0}\Delta _{0}\sim n$ (and
also for \textit{small detunings }$T_{0}\Delta _{0}\ll n$), the dynamics is
adiabatic for $\tau \sim 1$ only when $T_{0}\Omega _{0}\gg n.$

We can calculate for a power law coupling (\ref{Rabi0}) the scaled time $%
\tau _{M}=t_{M}/T_{0}$ at which the nonadiabatic coefficient $\widetilde{%
\gamma }(\tau )$ is maximum:%
\begin{equation}
\tau _{M}=\left( \frac{n-1}{2n+1}\right) ^{1/2n}\left( \frac{\Delta _{0}}{%
\Omega _{0}}\right) ^{1/n},
\end{equation}%
which gives the estimates%
\begin{equation}
\left\{
\begin{array}{c}
\tau _{M}=0,\text{ for }n=1, \\
\tau _{M}\sim \left( \frac{\Delta _{0}}{\Omega _{0}}\right) ^{1/n},\text{
for }n\geq 2%
\end{array}%
\right.
\end{equation}%
and%
\begin{equation}
\left\{
\begin{array}{c}
\widetilde{\gamma }(\tau _{M}=0)=\frac{1}{2}\frac{\Omega _{0}}{\Delta _{0}}%
\frac{1}{T_{0}\Delta _{0}},\text{ for }n=1, \\
\widetilde{\gamma }(\tau _{M})\sim \frac{n}{4}\left( \frac{\Omega _{0}}{%
\Delta _{0}}\right) ^{1/n}\frac{1}{T_{0}\Delta _{0}},\text{ for }n\geq 2.%
\end{array}%
\right.  \label{adiab param}
\end{equation}%
For $n=1$, the nonadiabatic coefficient $\widetilde{\gamma }(\tau )$
decreases monotonically from $\widetilde{\gamma }(\tau _{M}=0)=\frac{1}{2}%
\frac{\Omega _{0}}{\Delta _{0}}\frac{1}{T_{0}\Delta _{0}}$ to zero, with a
width of order $\frac{\Delta _{0}}{\Omega _{0}}$. For $n\geq 2$, it is
roughly bell-shaped and symmetric around $\tau _{M}$. This quantity (\ref%
{adiab param}) allows to characterize the \textit{global} adiabaticity: if $%
\widetilde{\gamma }(\tau _{M})\ll 1$, the dynamics is adiabatic at \textit{%
any time}. This implies that detunings such that $\Delta _{0}\gg \Omega _{0}$
induce adiabaticity at all times. For $\Delta _{0}\sim \Omega _{0}$, the
dynamics is also adiabatic at all times if the detuning is large $%
T_{0}\Delta _{0}\gg n$. (The case $\Delta _{0}\sim \Omega _{0}$ and $%
T_{0}\Delta _{0}\sim n$ is not of interest here since it induces a
nonadiabatic dynamics for $t\sim T_{0}$.)

Conversely detunings such that $\Delta _{0}\ll \Omega _{0}$ induce a
nonadiabatic dynamics around times $\tau _{M}$. This is this last non
trivial case, which is of interest here (accompanied with the condition $%
T_{0}\Omega _{0}\gg n$ to have adiabaticity beyond $\tau \sim 1$). This case
can be described in more detail as follows: during early scaled times of
order $\left( \frac{\Delta _{0}}{\Omega _{0}}\right) ^{1/n}$, the dynamics
is approximately \textit{adiabatic} for the rising coupling (this initial
adiabatic regime occurs only for $n>1$), it is followed by a \textit{%
nonadiabatic dynamics} around times $\tau _{M}$ which lasts during times
also of order $\left( \frac{\Delta _{0}}{\Omega _{0}}\right) ^{1/n}$ (true
for any $n\geq 1$), and an \textit{adiabatic evolution} for times beyond
(see Fig. \ref{Fig1_time}).

For exponential and Gaussian couplings, we will consider the non-trivial
case $\Delta_{0}\ll\Omega_{0}$ with $T_{0}\Omega_{0}\gg1$, which allows
adiabaticity [through Eq. (\ref{adiab_following})] for times $\tau\sim1$.

We will show the universality of this regime for the different couplings
considered above.

\subsection{Population history}

\begin{figure}[tb]
\centerline{\includegraphics[scale=0.75]{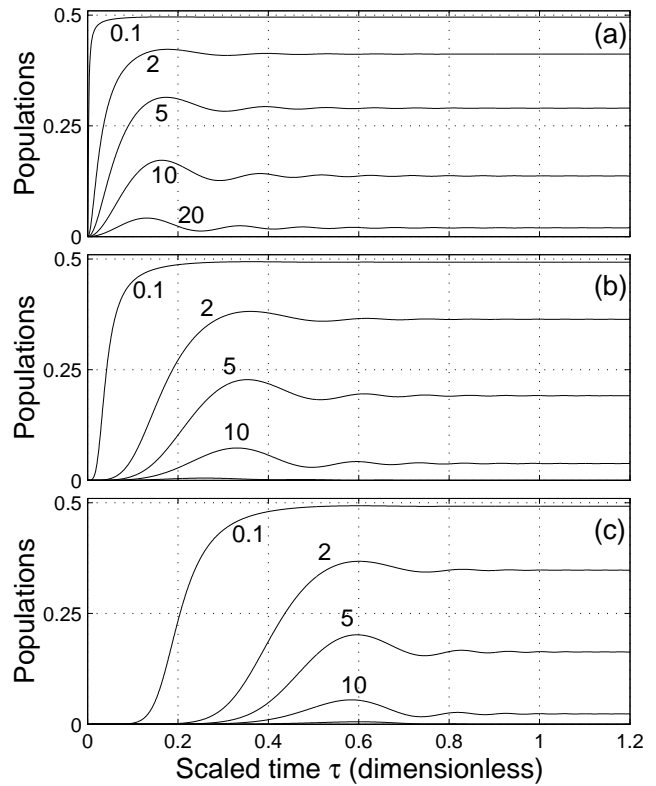}}
\caption{Adiabatic state population history $\left| A_{2}(\protect\tau %
)\right| ^{2}$ for the cases of linear rising coupling $n=1$ (frame a),
power law rising with $n=2$ (frame b) and $n=4$ (frame c) for different
dimensionless detuning $\Delta _{0}T_{0}$ (indicated close to each curve)
and a fixed coupling $\Omega _{0}T_{0}=100$ which allows the condition (\ref%
{ad2}) to be verified for the different cases considered. }
\label{Fig1_time}
\end{figure}

The time evolution of the population $\left| A_{+}(\tau )\right| ^{2}=\left|
\left\langle \Phi _{+}(\tau )\right| \left. \phi (\tau )\right\rangle
\right| ^{2}$ in the adiabatic state $\left| \Phi _{+}(\tau )\right\rangle $
is shown on the Fig. \ref{Fig1_time} for power law risings. One can see the
regions of the adiabatic and nonadiabatic evolution according to the
analysis given below. For $n=1$ the nonadiabatic evolution, characterized by
a change of the population which moves into $\left| \Phi _{+}(\tau
)\right\rangle $, starts already for short times and is followed by an
adiabatic evolution, characterized by a constant population in $\left| \Phi
_{+}(\tau )\right\rangle $ and $\left| \Phi _{-}(\tau )\right\rangle $. For $%
n>1$ there are time intervals of adiabacity for small and large times. For
higher $n$ the early time interval of adiabaticity is longer, since the
characteristic time of rising is larger. For larger $\Delta _{0}$ the early
time interval of adiabaticity is also longer.

\section{Lifting and creation of quasi-degeneracy by linearly rising coupling%
}

The problem of lifting of quasi-degeneracy can be solved analytically in the
case of linear rising of the coupling [Eq. (\ref{Rabi0}) with $n=1$]. The
exact solution of the Schr\"{o}dinger equation can be found in this case, in
terms of the parabolic cylinder functions (see Appendix \ref{linear}). We
need to calculate the dynamics at asymptotic times beyond the lifting of
quasi-degeneracy when the population of the adiabatic states is time
independent.

The initial condition $A_{-}(\tau _{i})=B_{-}(\tau _{i})=1,$ $A_{+}(\tau
_{i})=B_{+}(\tau _{i})=0$ leads to the amplitudes of the adiabatic states
\begin{subequations}
\label{transition}
\begin{eqnarray}
A_{\pm }(\tau ) &\rightsquigarrow &\frac{1}{\sqrt{2}}\left( a\mp be^{\mp
i\varphi }\right) e^{\mp i\left( \chi _{0}+\eta _{d}(\tau )\right) } \\
&\equiv &\sqrt{p_{\pm }}\ e^{\mp i\left( \chi _{\pm }+\eta _{d}(\tau
)\right) }
\end{eqnarray}%
with the transition probabilities
\end{subequations}
\begin{align}
p_{\pm }& =\frac{1}{2}\left| a\mp be^{i\varphi }\right| =\frac{1}{2}\left(
1\mp \sqrt{1-e^{-\pi \omega ^{2}}}\cos \varphi \right) ,  \label{pplus} \\
\varphi & =\arg \Gamma \left( 1-i\frac{\omega ^{2}}{4}\right) -\arg \Gamma
\left( \frac{1}{2}-i\frac{\omega ^{2}}{4}\right) +\frac{\pi }{4}, \\
\omega & =\frac{T_{0}\Delta _{0}}{\sqrt{2T_{0}\Omega _{0}}},  \label{omega}
\\
a& =\frac{1}{\sqrt{2}}\sqrt{1+e^{-\pi \omega ^{2}/2}},\ b=\frac{1}{\sqrt{2}}%
\sqrt{1-e^{-\pi \omega ^{2}/2}},
\end{align}%
(where $\Gamma $ denotes the Gamma--function) and the phases%
\begin{align}
\chi _{\pm }& =\chi _{0}+\arg \left( a\mp be^{i\varphi }\right) ,
\label{chiplus} \\
\chi _{0}& =\arg \Gamma \left( \frac{1}{2}-i\frac{\omega ^{2}}{4}\right) -%
\frac{\omega ^{2}}{4}\left( 1-\ln \frac{\omega ^{2}}{4}\right) , \\
\eta _{d}(\tau )& =\frac{1}{\hbar }\int_{0}^{\tau }\lambda _{+}(\tau )d\tau =%
\frac{T_{0}}{2}\int_{0}^{\tau }\sqrt{\Delta _{0}^{2}+\Omega _{0}^{2}\tau ^{2}%
}d\tau .  \label{etad}
\end{align}%
One can interpret $\sqrt{p_{\pm }}\exp \left( \pm i\chi _{\pm }\right) $ as
the probability amplitudes of the adiabatic states from the initial bare
state $\left| -\right\rangle $ resulting from the \textit{lifting of
degeneracy} and the\textit{\ splitting of the population}. This splitting is
accompanied by phases shifts $\pm \chi _{\pm }$. The additional phases $\pm
\eta _{d}(\tau )$ given by the time integral of the adiabatic eigenvalues
are thus the \textit{dynamical phases} of the process.

The accuracy of the asymptotics (\ref{transition}) is shown in Fig. \ref%
{Fig2}. At $\tau =1$, we observe already a precision of many digits both in
population and phase.

\begin{figure}[tb]
\centerline{\includegraphics[scale=0.75]{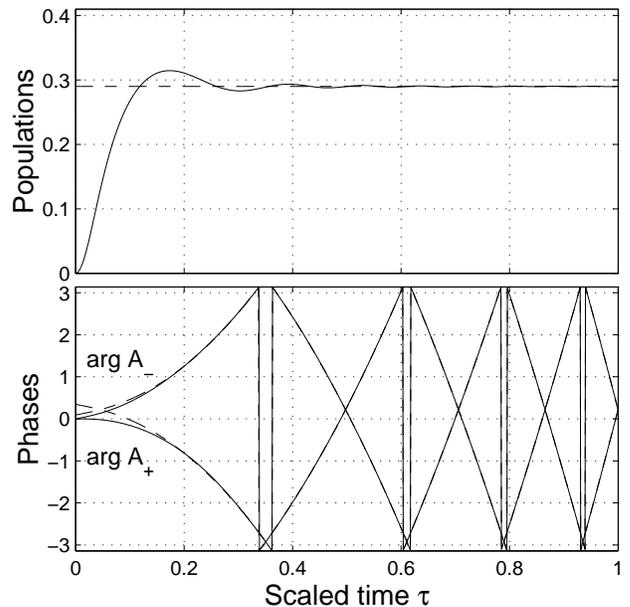}}
\caption{History of adiabatic state population $\left| A_{+}(\protect\tau%
)\right| ^{2}$ [upper frame; numerics: full line, and analytical formula $%
p_{+}$ (\ref{pplus}): dashed line] and phases $\arg A_{\pm}(\protect\tau)$
[lower frame; numerics: full line, and analytical formulae $\mp(\protect\chi%
_{\pm}+\protect\eta_{d}(\protect\tau))$ given by Eqs. (\ref{chiplus}) and (%
\ref{etad}): dashed lines] for the linear rising coupling with $%
T_{0}\Delta_{0}=5$ and $T_{0}\Omega_{0}=100$ (giving $\protect\omega\approx
0.35$). Phases are plotted in the interval $[-\protect\pi,+\protect\pi[$,
which induces artificial jumps that we have connected for a clearer
identification.}
\label{Fig2}
\end{figure}

\begin{figure}[tb]
\centerline{\includegraphics[scale=0.75]{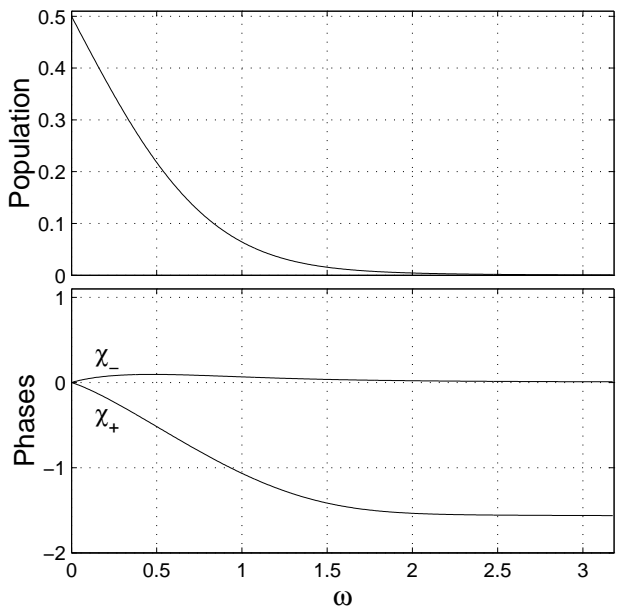}}
\caption{Asymptotic adiabatic state population $\left| A_{+}\right| ^{2}$ (%
\ref{pplus}) (upper frame) and phases $\protect\chi _{-}$ and $\protect\chi %
_{+}$ (\ref{chiplus}) (lower frame) as functions of the dimensionless
quantity $\protect\omega $ for the linear rising of the coupling.}
\label{Fig3}
\end{figure}

One essential result is that the adiabatic populations $p_{\pm }$ and the
phases $\chi _{\pm }$ \textit{depend only on} $\omega $ [Eq. (\ref{omega})].
Their dependence is shown on Fig. \ref{Fig3}. The phases $\chi _{+}$ and $%
\chi _{-}$ go asymptotically to $-\pi /2$ and $0$ respectively. One can
remark that $\chi _{-}$ is not very different from zero after (and also
during) the lifting of degeneracy for any detuning. This trend has been
numerically checked to occur for any $n$.

The reversed problem of lifting of quasi-degeneracy, which we called
creation of quasi-degeneracy, is induced by a pulse falling to zero [Eq. (%
\ref{Rabi0}), with $n=1$]. It leads to a recombination of the two adiabatic
states. This has been calculated in Appendix \ref{linear}.

\section{Lifting and creation of quasi-degeneracy by exponentially rising
coupling}

The problem of lifting and creation of quasi-degeneracy can be also solved
analytically in the case of exponentially rising (\ref{exp_rising}) and
falling (\ref{exp_falling}) coupling (see Appendix B). The asymptotics of
the exact solution can be expressed in terms of the Kummer functions. In the
adiabaticity region where the population of the eigenstates is time
independent, with the initial condition $A_{-}(\tau _{i})=B_{-}(\tau
_{i})=1, $ $A_{+}(\tau _{i})=B_{+}(\tau _{i})=0$ at $\tau _{i}\rightarrow
-\infty $, we obtain the amplitudes of the adiabatic states%
\begin{equation}
A_{\pm }(\tau )\rightsquigarrow \sqrt{p_{\pm }}\ e^{i\left( \xi \mp \zeta
(\tau )\right) }  \label{transexp}
\end{equation}%
with the transition probabilities%
\begin{equation}
p_{-}=\frac{1}{1+e^{-\pi \varpi }},\quad p_{+}=\frac{e^{-\pi \varpi }}{%
1+e^{-\pi \varpi }},  \label{pplusexp}
\end{equation}%
the instantaneous dimensionless pulse half-area (which is in fact an
instantaneous Rabi frequency half-area)%
\begin{equation}
\zeta (\tau )=\frac{T_{0}}{2}\int_{-\infty }^{\tau }\Omega (\tau ^{\prime
})d\tau ^{\prime }=\frac{T_{0}}{2}\Omega _{0}e^{\tau },  \label{sexp}
\end{equation}%
the dimensionless detuning%
\begin{equation}
\varpi =T_{0}\Delta _{0}
\end{equation}%
and the phase [given by Eq. (\ref{phase_ksi})]%
\begin{equation}
\xi =\arg \Gamma \left( \frac{1}{2}+i\frac{\varpi }{2}\right) +\varpi \ln 2-%
\frac{\varpi }{2}\ln 2\zeta (\tau _{i}).  \label{xi}
\end{equation}%
(In practice, $\tau _{i}$ is to be taken as a finite large negative number.)
The phase $\xi $ of the amplitudes is a \textit{common} phase for the
resulting superposition of adiabatic states. There is no additional relative
phase shift during the lifting of degeneracy.

It is remarkable that the transition probabilities \textit{depend only on
the detuning }$\Delta_{0}T_{0}$ (and not on $\Omega_{0}$). Moreover the
preceding dynamical phase is here replaced by a pulse area.

\begin{figure}[tb]
\centerline{\includegraphics[scale=0.75]{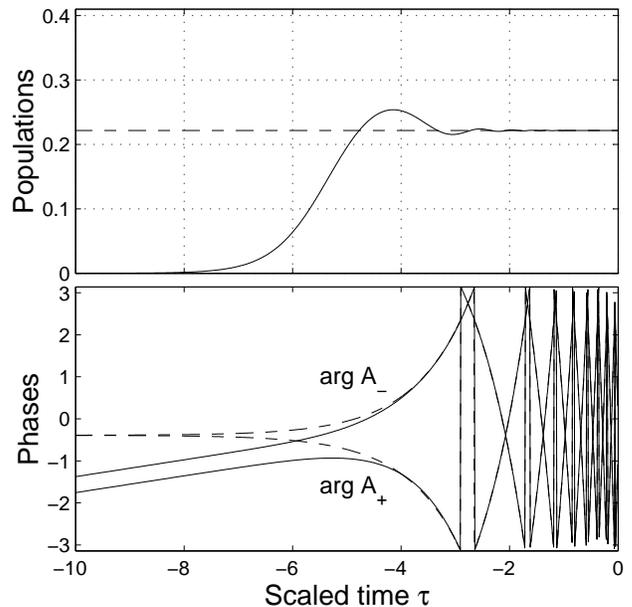}}
\caption{History of adiabatic state population $p_+=\left| A_{+}(\protect\tau%
)\right| ^{2}$ [upper frame; numerics: full line, and analytical formula $%
p_{+}$ (\ref{pplusexp}): dashed line] and phases $\arg A_{\pm}(\protect\tau)$
[lower frame; numerics: full line, and analytical formulae $\protect\xi\mp
\protect\zeta(\protect\tau)$ given by Eqs. (\ref{xi}) and (\ref{sexp}):
dashed lines] for the exponential rising of the coupling with $%
T_{0}\Delta_{0}=0.4$ and $T_{0}\Omega_{0}=100$.}
\label{Fig4}
\end{figure}

The accuracy of the asymptotics (\ref{transexp}) is shown in Fig. \ref{Fig4}.

\section{Perturbation theory for large detuning}

For large detuning $T_{0}\Delta _{0}$, the evolution of the system is
adiabatic at all times (we exclude $\Omega _{0}\gg \Delta _{0}$). We are
thus interested here in small nonadiabatic corrections from the initial
condition $A_{-}(\tau _{i})=B_{-}(\tau _{i})=1,$ $A_{+}(\tau
_{i})=B_{+}(\tau _{i})=0$. It is well known that their main contribution are
given by the nonsmoothness of the coupling at the beginning, characterized
here by a discontinuous $n^{\text{th}}$ derivative \cite{Garrido, Sancho}.
In this section we give estimates of corrections beyond.

It is convenient to use the adiabatic basis, in which we use the standard
perturbation theory:%
\begin{equation}
A_{\pm }(\tau )=A_{\pm }^{\left( 0\right) }(\tau )+\varepsilon A_{\pm
}^{\left( 1\right) }(\tau )+\varepsilon ^{2}A_{\pm }^{\left( 2\right) }(\tau
)+\cdots
\end{equation}%
with $\varepsilon \equiv 1/\left( T_{0}\left| \Delta _{0}\right| \right) $
and the initial condition $A_{-}^{\left( 0\right) }(\tau _{i})=1$, $%
A_{+}^{\left( 0\right) }(\tau _{i})=0$, $A_{\pm }^{\left( n>0\right) }(\tau
)=0$. The zeroth order gives
\begin{equation}
A_{-}^{\left( 0\right) }(\tau )=e^{i\eta _{d}(\tau )},\qquad A_{+}^{\left(
0\right) }(\tau )=0
\end{equation}%
with the dynamical phase%
\begin{equation}
\eta _{d}(\tau )=\frac{1}{\hbar }\int_{0}^{\tau }\lambda _{+}(\tau )d\tau =%
\frac{T_{0}}{2}\int_{\tau _{i}}^{\tau }\delta (\tau ^{\prime })d\tau
^{\prime }.
\end{equation}%
The first order contributions read
\begin{subequations}
\begin{align}
\varepsilon A_{-}^{\left( 1\right) }(\tau )& =0 \\
\varepsilon A_{+}^{\left( 1\right) }(\tau )& =\frac{1}{2}e^{-i\eta _{d}(\tau
)}\int_{\tau _{i}}^{\tau }d\tau ^{\prime }\gamma (\tau ^{\prime })e^{2i\eta
_{d}(\tau ^{\prime })},
\end{align}%
which give at first order
\end{subequations}
\begin{subequations}
\label{Alarge}
\begin{align}
A_{-}(\tau )& \approx e^{i\eta _{d}(\tau )} \\
A_{+}(\tau )& \approx \frac{1}{2}e^{-i\eta _{d}(\tau )}\int_{\tau
_{i}}^{\tau }d\tau ^{\prime }\gamma (\tau ^{\prime })e^{2i\eta _{d}(\tau
^{\prime })}.
\end{align}%
This shows that the phase of $A_{-}(\tau )$ is approximately given by only
the dynamical phase. This is consistent with what we obtained for $n=1$,
where $\chi _{-}$ (\ref{chiplus}) was not very different from zero during
the lifting of degeneracy. The transition probability at the first order $%
P_{+}(\tau )\equiv \left| A_{+}(\tau )\right| ^{2}$ at large times $\tau
\rightarrow +\infty $ is given asymptotically by
\end{subequations}
\begin{equation}
P_{+}(\tau )\rightsquigarrow \frac{1}{4}\left| \int_{\tau _{i}}^{\tau }d\tau
\ \gamma (\tau )e^{2i\eta _{d}(\tau )}\right| ^{2}.  \label{p+1}
\end{equation}%
For consistency of the perturbation theory, we should keep only terms of
order $\varepsilon $ of the integral by partial integration, using $\exp
\left( 2i\eta _{d}(\tau )\right) =\exp \left[ \frac{i}{\varepsilon }%
\int_{\tau _{i}}^{\tau }\sqrt{1+\left( \Omega /\Delta _{0}\right) ^{2}}d\tau
^{\prime }\right] $. We however keep here the full expression (\ref{p+1}) as
is usually done \cite{Davis}.

We now consider the power rising coupling [Eq. (\ref{Rabi0})] with arbitrary
$n$. Introducing the large dimensionless parameter%
\begin{equation}
\alpha _{n}\equiv T_{0}\Delta _{0}\left( \frac{\Delta _{0}}{\Omega _{0}}%
\right) ^{1/n}\gg 1
\end{equation}%
and using $x=\left( \Omega _{0}/\Delta _{0}\right) ^{1/n}\tau $, one obtains%
\begin{equation}
\varepsilon A_{+}^{\left( 1\right) }(\tau )=J_{n}e^{-i\eta _{d}(\tau )}
\end{equation}%
with
\begin{subequations}
\begin{align}
J_{n}& \equiv \frac{1}{2}\int_{0}^{\infty }dx\frac{nx^{n-1}}{1+x^{2n}}%
e^{i\alpha _{n}\int_{0}^{x}\sqrt{1+u^{2n}}}du \\
& =\int_{0}^{\infty }d\omega \ g_{n}(\omega )e^{i\alpha _{n}\omega },
\end{align}%
defining as usually done
\end{subequations}
\begin{equation}
\omega (x)=\int_{0}^{x}\sqrt{1+u^{2n}}du
\end{equation}%
and%
\begin{equation}
g_{n}(\omega )=\frac{f_{n}\left[ x(\omega )\right] )}{\sqrt{1+\left[
x(\omega )\right] ^{2n}}},\quad f_{n}(x)=\frac{1}{2}\frac{nx^{n-1}}{1+x^{2n}}%
.
\end{equation}%
The probability to the first order reads
\begin{equation}
P_{+}(\infty )\approx \left| J_{n}\right| ^{2}.
\end{equation}%
We are interested in the asymptotics of $J_{n}$ for $\alpha _{n}\gg 1.$ One
can remark that the lower limit of integration in $J_{n}$ is $x_{\text{low}%
}=0$, unlike the more standard case where $x_{\text{low}}\rightarrow -\infty
$ \cite{Davis}$.$ This difference leads to an additional nonexponential
contribution of the integral $J$, dominant for $\alpha _{n}\gg 1$, which can
be calculated by $n$ partial integrations:%
\begin{equation}
J_{n}=S_{n}+I_{n}
\end{equation}%
with%
\begin{equation}
S_{n}\equiv \sum_{k=1}^{n}\left( \frac{i}{\alpha _{n}}\right) ^{k}\partial
_{\omega }^{k-1}g_{n}(0)
\end{equation}%
and%
\begin{equation}
I_{n}\equiv \left( \frac{-1}{i\alpha _{n}}\right) ^{n}\int_{0}^{\infty
}\partial _{\omega }^{n}g_{n}(\omega )e^{i\alpha _{n}\omega },  \label{Inint}
\end{equation}%
since $g_{n}(\infty )=0$. We calculate $\partial _{\omega
}^{k-1}g_{n}(0)=\partial _{x}^{k-1}f_{n}(0)=\delta _{k,n}\frac{1}{2}n!$ for $%
k\leq n$, giving for the nonexponential contribution%
\begin{equation}
S_{n}=\frac{1}{2}\left( \frac{i}{\alpha _{n}}\right) ^{n}n!  \label{Sn}
\end{equation}%
The asymptotics for $\alpha _{n}\rightarrow \infty $ of the exponential
contribution can be estimated as%
\begin{equation}
I_{n}\rightsquigarrow \left\{
\begin{array}{c}
\frac{\pi }{3}\sum_{k=0}^{n/2-1}F_{k}\text{,\quad for even }n \\
\frac{\pi }{3}\left[ \frac{1}{2}(-1)^{\frac{n-1}{2}}e^{-\alpha
_{n}b_{n}}+\sum_{k=0}^{\frac{n-3}{2}}F_{k}\right] \text{,\ for odd }n%
\end{array}%
\right.  \label{In}
\end{equation}%
with
\begin{equation}
F_{k}=(-1)^{k}e^{-\alpha _{n}b_{n}\sin \left[ \frac{\pi }{n}\left( k+\frac{1%
}{2}\right) \right] }e^{i\alpha _{n}b_{n}\cos \left[ \frac{\pi }{n}\left( k+%
\frac{1}{2}\right) \right] }
\end{equation}%
and%
\begin{equation}
b_{n}\equiv \int_{0}^{1}\sqrt{1-x^{2n}}dx=\frac{1}{4n}\sqrt{\pi }\frac{%
\Gamma \left( \frac{1}{2n}\right) }{\Gamma \left( \frac{3n+1}{2n}\right) }.
\end{equation}%
We can remark that the prefactor $\pi /3$ \ in the exponential contribution $%
I_{n}$ is wrong. The comparison with the exact result (\ref{pplus}) for $n=1$
gives\ the correct prefactor which should be $1$. This well known ''$\pi /3-$%
problem''\ is due to the fact that one has considered the first-order
perturbation theory in the adiabatic basis \cite{Davis}.

\begin{figure}[tb]
\centerline{\includegraphics[scale=0.75]{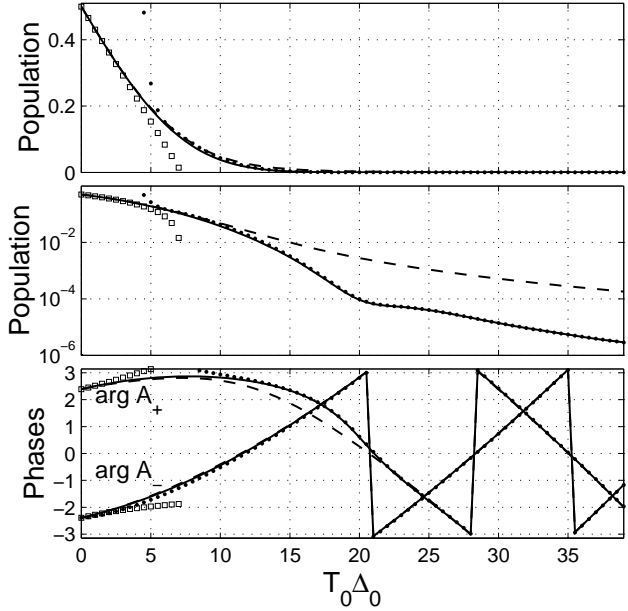}}
\caption{Asymptotic adiabatic state population $P_{+}=\left| A_{+}\right|
^{2}$ [upper frame; numerics (full lines), perturbation theory for large
detuning (\ref{p+1}) (dotted lines), for small detuning (\ref{Pplus_small})
(squared lines) and approximate formula (\ref{pplus_any})], the same with
semi-log scale plot (middle frame), and phases $\arg A_{-}$ and $\arg A_{+}$
[lower frame; numerics (full lines), using approximations (\ref{Alarge})
(dotted lines), (\ref{Aplus_small}) (squared lines) and (\ref{Aplus_any})]
as functions of $T_{0}\Delta _{0}$ for the rising of the coupling with $n=2$
and $T_{0}\Omega_{0}=100$ (calculated at the asymptotic time $\protect\tau %
=3T_{0}$).}
\label{Fig5}
\end{figure}

Figure \ref{Fig5} shows the amplitudes for $n=2$ obtained with numerics
(full lines) and with the preceding formulae (dotted lines). The agreement
is good from $T_{0}\Delta_{0}\approx5$ for the probability and from $%
T_{0}\Delta _{0}\approx10$ for the phase $\arg A_{+}$. The agreement is good
for $\arg A_{-}$ for any $\omega$. We have indeed found numerically $\arg
A_{-}\approx\eta_{d}$ for any $\omega$ during the lifting of degeneracy.

\begin{figure}[tb]
\centerline{\includegraphics[scale=0.75]{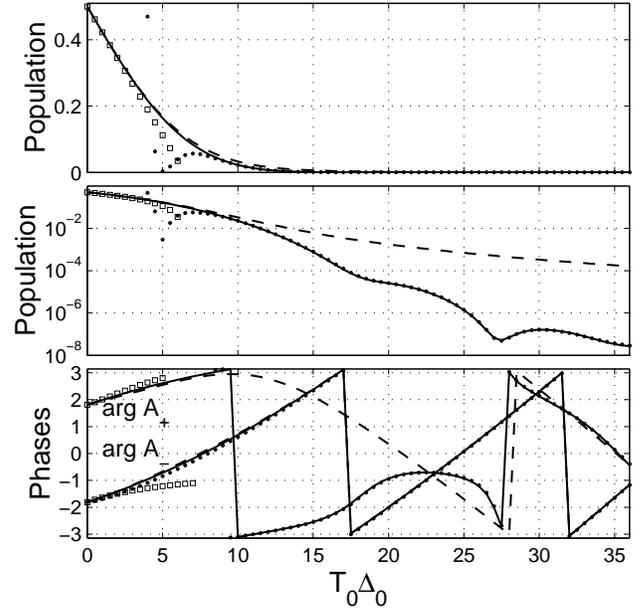}}
\caption{Same as Fig. 5 for the rising of the coupling with $n=3$
(calculated at $\protect\tau =2.5T_{0}$).}
\label{Fig6}
\end{figure}
For a smooth lifting of degeneracy [with a coupling such as exponential (\ref%
{exp_rising}) or Gaussian coupling (\ref{Gaussian})], the terms of type $S$ (%
\ref{Sn}) are all zero. Only an exponentially small term contributes to the
lifting of quasi-degeneracy for large detuning.

\section{Perturbation theory for small detuning}

For small detuning $T_{0}\Delta _{0}\ll 1$ it is convenient to apply
perturbation theory with $\varepsilon =T_{0}\Delta _{0}$ in the Landau-Zener
basis as described below. The amplitudes $\phi _{LZ}(\tau )$ of the states
in the Landau-Zener basis are obtained from the amplitudes of the bare
states via the time-independent unitary transformation $\mathsf{S}$:%
\begin{equation}
\phi _{LZ}(\tau )\equiv \left[
\begin{array}{c}
Z_{-}(\tau ) \\
Z_{+}(\tau )%
\end{array}%
\right] =\mathsf{S}^{\dagger }\phi (\tau ),
\end{equation}%
with
\begin{equation}
\mathsf{S}=\frac{1}{\sqrt{2}}\left[
\begin{array}{cc}
1 & 1 \\
-1 & 1%
\end{array}%
\right] ,
\end{equation}%
giving the Landau-Zener Hamiltonian
\begin{subequations}
\begin{align}
\mathsf{H}_{LZ}(\tau )& \equiv \mathsf{S}^{\dagger }\mathsf{H}(\tau )\mathsf{%
S} \\
& =\frac{\hbar }{2}\left[
\begin{array}{cc}
-\Omega (\tau ) & -\Delta _{0} \\
-\Delta _{0} & \Omega (\tau )%
\end{array}%
\right] ,
\end{align}%
associated to the Schr\"{o}dinger equation
\end{subequations}
\begin{equation}
i\hbar \frac{\partial \phi _{LZ}}{\partial \tau }(\tau )=T_{0}\mathsf{H}%
_{LZ}(\tau )\phi _{LZ}(\tau ),
\end{equation}%
and the initial conditions
\begin{equation}
Z_{\pm }(\tau _{i})=\frac{1}{\sqrt{2}},
\end{equation}%
since one starts with $A_{-}(\tau _{i})=B_{-}(\tau _{i})=1,$ $A_{+}(\tau
_{i})=B_{+}(\tau _{i})=0$ [see Eq. (\ref{phiA})]. The zero order of
perturbation theory gives
\begin{equation}
Z_{\pm }^{(0)}(\tau )=\frac{1}{\sqrt{2}}e^{\mp i\zeta (\tau )}
\end{equation}%
where $\zeta (\tau )$ is the instantaneous dimensionless pulse half-area%
\begin{equation}
\zeta (\tau )=\frac{T_{0}}{2}\int_{\tau _{i}}^{\tau }\Omega (\tau ^{\prime
})d\tau ^{\prime }.
\end{equation}%
The first order contribution reads
\begin{equation}
\varepsilon Z_{\pm }^{(1)}(\tau )=i\frac{T_{0}\Delta _{0}}{2\sqrt{2}}e^{\mp
i\zeta (\tau )}\int_{\tau _{i}}^{\tau }d\tau ^{\prime }e^{\pm 2i\zeta (\tau
^{\prime })},
\end{equation}%
leading to the first order solution%
\begin{equation}
Z_{\pm }(\tau )\approx Z_{\pm }^{(0)}(\tau )+\varepsilon Z_{\pm }^{(1)}(\tau
).
\end{equation}%
For large times $\tau \rightarrow +\infty $, $\mathsf{H}_{LZ}(\tau )$
becomes diagonal and $\phi _{LZ}(\tau )$ asymptotically coincides with $\phi
_{A}(\tau )$, which gives at first order:%
\begin{equation}
A_{\pm }(\tau )\rightsquigarrow \frac{1}{\sqrt{2}}e^{\mp i\zeta (\tau
)}\left( 1+i\frac{T_{0}\Delta _{0}}{2}\int_{\tau _{i}}^{\tau }d\tau ^{\prime
}e^{\pm 2i\zeta (\tau ^{\prime })}\right) .
\end{equation}%
The transition probability at first order can be thus written as
\begin{subequations}
\begin{equation}
P_{+}(\tau )\rightsquigarrow \frac{1}{2}\left[ 1-T_{0}\Delta _{0}\int_{\tau
_{i}}^{\tau }d\tau ^{\prime }\sin \left( 2\zeta (\tau ^{\prime })\right) %
\right] .
\end{equation}%
For the power law rising coupling, one thus obtains for large times $\tau
\rightarrow +\infty $%
\end{subequations}
\begin{equation}
A_{\pm }(\infty )\approx \frac{1}{\sqrt{2}}e^{\mp i\zeta (\tau )}\left[ 1+%
\frac{T_{0}\Delta _{0}}{2}\left( \mp K_{n}+iL_{n}\right) \right]
\label{Aplus_small}
\end{equation}%
and%
\begin{equation}
P_{+}(\infty )\approx \frac{1}{2}\left( 1-T_{0}\Delta _{0}K_{n}\right)
\label{Pplus_small}
\end{equation}%
with
\begin{equation}
\zeta (\tau )=\frac{T_{0}\Omega _{0}}{2}\frac{\tau ^{n+1}}{n+1},
\end{equation}%
\begin{subequations}
\begin{align}
K_{n}& =\left( \frac{n+1}{T_{0}\Omega _{0}}\right) ^{\frac{1}{n+1}%
}\int_{0}^{\infty }ds\sin \left( s^{n+1}\right) \\
& =\left( \frac{1}{T_{0}\Omega _{0}}\right) ^{\frac{1}{n+1}}\frac{\sqrt{\pi }%
}{\left[ 2\left( n+1\right) \right] ^{n/\left( n+1\right) }}\frac{\Gamma
\left( \frac{n+2}{2\left( n+1\right) }\right) }{\Gamma \left( \frac{2n+1}{%
2\left( n+1\right) }\right) }
\end{align}%
and
\end{subequations}
\begin{subequations}
\begin{align}
L_{n}& =\left( \frac{n+1}{\Omega _{0}T_{0}}\right) ^{\frac{1}{n+1}%
}\int_{0}^{\infty }ds\cos \left( s^{n+1}\right) \\
& =\left( \frac{1}{T_{0}\Omega _{0}}\right) ^{\frac{1}{n+1}}\frac{\sqrt{\pi }%
}{\left[ 2\left( n+1\right) \right] ^{n/\left( n+1\right) }}\frac{\Gamma
\left( \frac{1}{2\left( n+1\right) }\right) }{\Gamma \left( \frac{n}{2\left(
n+1\right) }\right) } \\
& =\frac{\sin \left( \frac{\pi }{2}\frac{n+2}{n+1}\right) }{\sin \left(
\frac{\pi }{2}\frac{2n+1}{n+1}\right) }K_{n}.
\end{align}%
For $n=1$, one has in particular $K_{1}=\sqrt{\pi /T_{0}\Omega _{0}}/2=\sqrt{%
\pi /2}\omega /T_{0}\Delta _{0}$ and
\end{subequations}
\begin{equation}
P_{+}(\infty )\approx \frac{1}{2}\left( 1-\sqrt{\frac{\pi }{2}}\omega
\right) ,
\end{equation}%
with $\omega $ given by Eq. (\ref{omega}), which is recovered from the exact
asymptotic solution (\ref{pplus}) in the limit of small detuning.

Figure \ref{Fig5} shows the transition probability and the phases for $n=2$
obtained with the preceding formulas. The agreement is good until $%
T_{0}\Delta _{0}\approx 5$. The two perturbative formulas for small and
large detunings almost match for the probability. Only a small region of
intermediate detuning is not covered by either approximation.

\begin{figure}[tb]
\centerline{\includegraphics[scale=0.75]{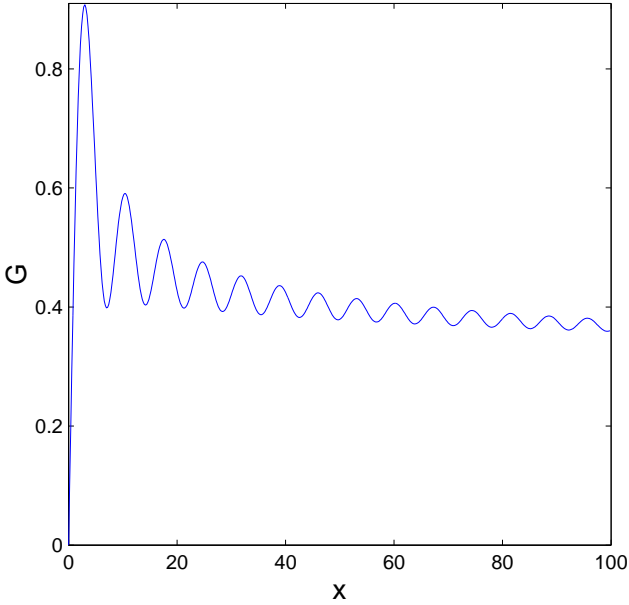}}
\caption{Function $G(x)$ as defined in Eq. (\ref{Gx}) involved in
the population transfer (\ref{PplusG}) for the Gaussian coupling
and small detunings.} \label{gauss}
\end{figure}

In the case of the exponential coupling (\ref{exp_rising}) with $\tau
_{i}\rightarrow-\infty$, we obtain, using the variable $x=T_{0}\Delta_{0}%
\exp(\tau)$, in the asymptotic limit $\tau\rightarrow\infty$:
\begin{subequations}
\begin{align}
P_{+}(\infty) & \approx\frac{1}{2}\left( 1-T_{0}\Delta_{0}\int_{0}^{\infty
}dx\frac{\sin x}{x}\right) \\
& =\frac{1}{2}\left( 1-\frac{\pi}{2}T_{0}\Delta_{0}\right) ,
\end{align}
which coincides with the exact asymptotic result (\ref{pplusexp}) in the
limit of small detuning.

For the Gaussian coupling (\ref{Gaussian}), one obtains with the asymptotic
limit taken at $\tau =0$:
\end{subequations}
\begin{equation}
P_{+}(0)=\frac{1}{2}\left[ 1-G(T_{0}\Omega _{0})T_{0}\Delta _{0}\right]
\label{PplusG}
\end{equation}%
with%
\begin{equation}
G(x)=\int_{-\infty }^{0}d\tau \sin \left[ \frac{\sqrt{\pi }}{2}x\left( 1+%
\text{erf}\left( \tau \right) \right) \right] .  \label{Gx}
\end{equation}%
The function $G(x)$ is shown in Fig. \ref{gauss}. It oscillates less and
less for larger $x$, and goes very slowly to 0. We are interested in the
adiabatic limit $x\equiv T_{0}\Omega _{0}\gg 1$, where $G(x)$ depends weakly
on $x$. This allows us to conclude that, in the limit of small detunings and
large $T_{0}\Omega _{0}$, the population transfer is quite robust with
respect to the peak amplitude: $\lim_{T_{0}\Omega _{0}\gg 1}\left|
dP_{+}(0)/d\left( T_{0}\Omega _{0}\right) \right| \ll 1$. Already for $%
T_{0}\Omega _{0}\sim 10$, we have $\left| dP_{+}(0)/d\left( T_{0}\Omega
_{0}\right) \right| <0.05\times T_{0}\Delta _{0}$. Moreover we can remark
that it is a bit more robust with respect to the detuning than the
exponentially rising case: for $T_{0}\Omega _{0}\sim 10$, we have $\left|
dP_{+}(0)/d\left( T_{0}\Delta _{0}\right) \right| <0.3$ (to be compared with
the exponentially rising, where we have the constant value $\left|
dP_{+}(0)/dT_{0}\Delta _{0}\right| =\pi /4\approx 0.79$).

\section{Approximate formula for arbitrary detuning}

The result (\ref{Pplus_small}) for small detuning suggests to extend for
arbitrary $n$ the exact results of $n=1$ [Eqns (\ref{transition}) to (\ref%
{etad})] replacing in these formulae $\omega $ by $\omega _{n}$:
\begin{subequations}
\label{omegan}
\begin{align}
\omega _{n}& \equiv \sqrt{\frac{2}{\pi }}K_{n}T_{0}\Delta _{0}
\label{omeganK} \\
& =T_{0}\Delta _{0}\left( \frac{1}{T_{0}\Omega _{0}}\right) ^{\frac{1}{n+1}}%
\frac{\sqrt{2}}{\left[ 2(n+1)\right] ^{n/\left( n+1\right) }}\frac{\Gamma
\left( \frac{n+2}{2\left( n+1\right) }\right) }{\Gamma \left( \frac{2n+1}{%
2\left( n+1\right) }\right) }.
\end{align}%
Asymptotic analysis allows us to extend the phases as follows: For large $%
T_{0}\Delta _{0}$, we take into account the $S_{n}$ contribution (\ref{Sn})
[which gives the additional prefactor $n$ in Eq. (\ref{Chinplus})]; for
small $T_{0}\Delta _{0}$, we have used (\ref{Aplus_small}) [which gives the
additional sine ratio in Eq. (\ref{Chinmoins})]. We finally obtain
\end{subequations}
\begin{equation}
A_{n,\pm }(\tau )\rightsquigarrow \sqrt{p_{n,\pm }}\ e^{\mp i\left( \chi
_{n,\pm }+\eta _{n}(\tau )\right) }  \label{Aplus_any}
\end{equation}%
with the transition probabilities%
\begin{align}
p_{n,\pm }& \equiv \left| A_{n,\pm }\right| ^{2}=\frac{1}{2}\left( 1\mp
\sqrt{1-e^{-\pi \omega _{n}^{2}}}\cos \varphi _{n}\right) ,
\label{pplus_any} \\
\varphi _{n}& =\arg \Gamma \left( 1-i\frac{\omega _{n}^{2}}{4}\right) -\arg
\Gamma \left( \frac{1}{2}-i\frac{\omega _{n}^{2}}{4}\right) +\frac{\pi }{4},
\end{align}%
and the phases%
\begin{align}
\chi _{n,-}& =\frac{\sin \left( \frac{\pi }{2}\frac{2n+1}{n+1}\right) }{\sin
\left( \frac{\pi }{2}\frac{n+2}{n+1}\right) }\chi _{n,0}+\arg \left(
a_{n}+b_{n}e^{i\varphi _{n}}\right) ,  \label{Chinmoins} \\
\chi _{n,+}& =\chi _{n,0}+n\arg \left( a_{n}-b_{n}e^{i\varphi _{n}}\right) ,
\label{Chinplus} \\
\chi _{n,0}& =\arg \Gamma \left( \frac{1}{2}-i\frac{\omega _{n}^{2}}{4}%
\right) -\frac{\omega _{n}^{2}}{4}\left( 1-\ln \frac{\omega _{n}^{2}}{4}%
\right) , \\
a_{n}& =\frac{1}{\sqrt{2}}\sqrt{1+e^{-\pi \omega _{n}^{2}/2}},\ b_{n}=\frac{1%
}{\sqrt{2}}\sqrt{1-e^{-\pi \omega _{n}^{2}/2}}, \\
\eta _{n}(\tau )& =\frac{T_{0}}{2}\int_{0}^{\tau }\sqrt{\Delta
_{0}^{2}+\Omega _{0}^{2}\tau ^{2n}}d\tau .
\end{align}%
It is remarkable that the amplitude depends only on $\omega _{n}$ (and also
on the dynamical phase $\eta _{n}(\tau )$ and $n$). This quantity gives the
scaling of the lifting of degeneracy. Figure \ref{Fig5} shows the good
accuracy of formula (\ref{Aplus_any}) for any detuning for $n=2$. Only a
logarithmic scale shows that the probablities are not as precise as
perturbation theory for large detuning. Figure \ref{Fig6} displays the
result for $n=3$. The populations $p_{2,\pm }$ and the phase $\arg A_{2,-}$
are quite good for any detuning. The phase $\arg A_{2,+}$ is well reproduced
for $T_{0}\Delta _{0}<10$.

One can remark that, as a function of the detuning, for $T_{0}\Delta _{0}<5$
, the phases change less fast than the populations. Thus the phases are more
robust than the populations with respect to the detuning..

\section{Application to the lineshape of the resonant excitation by strong
pulses}

In this section we apply the preceding formulation to calculate the
transition probability after a pulse excitation of bell-like shape and of
large area%
\begin{equation}
\mathcal{A}\equiv 2\zeta (\tau _{f})=T_{0}\int_{\tau _{i}}^{\tau _{f}}\Omega
(\tau ^{\prime })d\tau ^{\prime }\gg 1
\end{equation}%
to allow one to apply the preceding analysis for which a time where dynamics
is adiabatic must occur. We consider the secant hyperbolic pulse coupling
and recover the well-known Rosen-Zener formula. Discontinuous derivative
endings are next considered with the examples of truncated trig-pulses. We
study the lineshape, i.e. the transition probability as a function of the
detuning.

We consider the initial condition $B_{-}(\tau _{i})=1$, $B_{+}(\tau _{i})=0$.

\subsection{General calculation}

Between the lifting and creation of quasi-degeneracy, the pulse is assumed
to be an arbitrary smooth function between $\tau _{i}$ and $\tau _{f}$,
sufficiently far from zero to avoid intermediate creation of degeneracy. The
time evolution operator can thus be decomposed as:
\begin{subequations}
\begin{eqnarray}
U(\tau _{f},\tau _{i}) &=&\left[
\begin{array}{cc}
U_{11}(\tau _{f},\tau _{i}) & U_{12}(\tau _{f},\tau _{i}) \\
-\left[ U_{12}(\tau _{f},\tau _{i})\right] ^{\ast } & \left[ U_{11}(\tau
_{f},\tau _{i})\right] ^{\ast }%
\end{array}%
\right] \\
&=&\mathsf{R}(\tau _{f})U_{c}(\tau _{f},\tau _{2})U_{a}(\tau _{2},\tau
_{1})U_{\ell }(\tau _{1},\tau _{i})\mathsf{R}^{\dagger }(\tau _{i}),
\label{Upulse}
\end{eqnarray}%
where [see Eqs. (\ref{UA}), (\ref{UAplus})]
\end{subequations}
\begin{subequations}
\begin{equation}
U_{\ell }(\tau _{1},\tau _{i})=\left[
\begin{array}{cc}
U_{A+}^{11}(\tau _{1},\tau _{i}) & U_{A+}^{12}(\tau _{1},\tau _{i}) \\
-\left[ U_{A+}^{12}(\tau _{1},\tau _{i})\right] ^{\ast } & \left[
U_{A+}^{11}(\tau _{1},\tau _{i})\right] ^{\ast }%
\end{array}%
\right]
\end{equation}%
and [see Eq. (\ref{UAmoins_})]
\end{subequations}
\begin{subequations}
\begin{equation}
U_{c}(\tau _{f},\tau _{2})=\left[
\begin{array}{cc}
U_{A+}^{11}(\tau _{2}^{\prime },\tau _{f}) & -\left[ U_{A+}^{12}(\tau
_{2}^{\prime },\tau _{f})\right] ^{\ast } \\
U_{A+}^{12}(\tau _{2}^{\prime },\tau _{f}) & \left[ U_{A+}^{11}(\tau
_{2}^{\prime },\tau _{f})\right] ^{\ast }%
\end{array}%
\right] ,
\end{equation}%
with $\tau _{2}^{\prime }=-\tau _{2}+2\tau _{f}$ for the power law falling
and $\tau _{2}^{\prime }=-\tau _{2}$ for a smooth (exponential or Gaussian)
falling, are associated to the evolution operators of respectively lifting
and creation of degeneracy in the adiabatic basis, and
\end{subequations}
\begin{equation}
U_{a}(\tau _{2},\tau _{1})\mathbf{=}\left[
\begin{array}{cc}
e^{i\left[ \eta _{d}(\tau _{2})-\eta _{d}(\tau _{1})\right] } & 0 \\
0 & e^{-i\left[ \eta _{d}(\tau _{2})-\eta _{d}(\tau _{1})\right] }%
\end{array}%
\right] ,
\end{equation}%
with%
\begin{equation}
\eta _{d}(\tau )=\frac{T_{0}}{2}\int_{0}^{\tau }d\tau \sqrt{\Delta
_{0}^{2}+\Omega ^{2}(\tau )},  \label{etad2}
\end{equation}%
is associated to the adiabatic evolution between. Since the adiabatic states
coincide with the bare states at early and late times, we have $\mathsf{R}%
^{\dagger }(\tau _{i})=\mathsf{R}(\tau _{f})=\openone$. The amplitudes of
the bare states at the end of the pulse read thus
\begin{subequations}
\label{finalB}
\begin{eqnarray}
B_{-}(\tau _{f}) &=&U_{11}(\tau _{f},\tau _{i}), \\
B_{+}(\tau _{f}) &=&-\left[ U_{12}(\tau _{f},\tau _{i})\right] ^{\ast }.
\label{Bplus_final}
\end{eqnarray}

\subsection{Secant hyperbolic pulse}

The coupling reads in this case
\end{subequations}
\begin{equation*}
\Omega (\tau )=\Omega _{0}\sec \text{h}(\tau )\equiv \frac{\Omega _{0}}{%
\cosh (\tau )}
\end{equation*}%
with $\tau _{i}\rightarrow -\infty $ and $\tau _{f}\rightarrow +\infty $,
whose asymptotics rise and fall exponentially: $\Omega (\tau
)\rightsquigarrow 2\Omega _{0}\exp (\mp \tau )$ for $\tau \rightarrow \pm
\infty $. We thus calculate $B_{+}(\tau _{f})$ applying Eqs. (\ref%
{Bplus_final}) and (\ref{Upulse}), and using the asymptotic result [Eqs. (%
\ref{UA_exp})], which requires $s(\tau _{1})\equiv T_{0}\int_{-\infty
}^{\tau _{1}}\Omega (\tau )d\tau =T_{0}\Omega _{0}\exp (\tau _{1})\gg \varpi
\equiv T_{0}\Delta _{0}$ (with $\tau _{1}<0$) for the lifting of
quasi-degeneracy (see Appendix B) and $T_{0}\Omega _{0}\exp (-\tau _{2})\gg
T_{0}\Delta _{0}$ (with $\tau _{2}>0$) for the creation of quasi-degeneracy.
These two conditions are satisified for $\Omega _{0}\gg \Delta _{0}$.

In this limit, we obtain
\begin{subequations}
\begin{eqnarray}
B_{+}(\infty ) &\rightsquigarrow &-i\frac{\sin \left[ T_{0}\int_{-\infty
}^{+\infty }\Omega (\tau )d\tau /2\right] }{\cosh \left( \pi T_{0}\Delta
_{0}/2\right) } \\
&=&-i\frac{\sin \left( \pi T_{0}\Omega _{0}/2\right) }{\cosh \left( \pi
T_{0}\Delta _{0}/2\right) }, \\
B_{-}(\infty ) &\rightsquigarrow &e^{2i\xi }[\cos \left( \pi T_{0}\Omega
_{0}/2\right)  \notag \\
&&+i\sin \left( \pi T_{0}\Omega _{0}/2\right) \tanh (\pi T_{0}\Delta _{0}/2)]
\end{eqnarray}%
with $\xi $ defined in (Eq. \ref{xi}), which allows to recover the
well-known Rosen-Zener formula \cite{RosenZener}
\end{subequations}
\begin{equation}
\left| B_{+}(\infty )\right| ^{2}=\frac{\sin ^{2}\left( T_{0}\Omega _{0}\pi
/2\right) }{\cosh ^{2}\left( \pi T_{0}\Delta _{0}/2\right) }\equiv P_{RZ},
\end{equation}%
which is exact for any $\Omega _{0}$ and $\Delta _{0}$. Note also that the
phase of $B_{+}(\infty )$ is also exact, but that the phase of $B_{-}(\infty
)$ is only approximatively valid for small $T_{0}\Delta _{0}$.

\subsection{Truncated pulses of linear endings}

We assume that the pulse starts and ends with linear rising and falling
around respectiveley $\tau _{i}=0$ and $\tau _{f}=\tau _{p}\gg 1:$%
\begin{subequations}
\begin{eqnarray}
\Omega (\tau ) &\sim &\Omega _{0}\tau ,\quad 1>\tau \geq 0, \\
\Omega (\tau ) &\sim &\Omega _{0}\left( \tau _{p}-\tau \right) ,\quad \tau
_{p}-1<\tau \leq \tau _{p}
\end{eqnarray}%
with the discontinuous first derivative
\end{subequations}
\begin{equation}
\Omega _{0}=\left. \frac{d\Omega }{d\tau }\right| _{\tau =0}=-\left. \frac{%
d\Omega }{d\tau }\right| _{\tau =\tau _{p}},
\end{equation}%
such that the lifting (resp. creation) of degeneracy occurs during the
linear rising (resp. falling) of the pulse. We assume $\tau _{1}\sim 1,$ $%
\tau _{2}\sim \tau _{f}-1$ in Eq. (\ref{Upulse}). We obtain the amplitudes
of the bare states at the end of the pulse
\begin{subequations}
\label{finalB_linear}
\begin{eqnarray}
B_{-}(\tau _{p}) &=&p_{-}e^{i\left( 2\chi _{-}+\eta _{d}(\tau _{p})\right)
}+p_{+}e^{-i\left( 2\chi _{+}+\eta _{d}(\tau _{p})\right) } \\
&=&\cos \psi \cos \varphi +\sin \psi \\
&&\times \left( i\sqrt{1-e^{-\pi \omega ^{2}}}+e^{-\pi \omega ^{2}/2}\sin
\varphi \right) \\
B_{+}(\tau _{p}) &=&-2i\sqrt{p_{+}p_{-}}\sin \left( \chi _{-}+\chi _{+}+\eta
_{d}(\tau _{p})\right) \\
&=&i\left( \cos \psi \sin \varphi -e^{-\pi \omega ^{2}/2}\sin \psi \cos
\varphi \right)  \label{finalBplus}
\end{eqnarray}%
with $p_{\pm }$ and $\chi _{\pm }$ respectively defined in Eqs. (\ref{pplus}%
) and (\ref{chiplus}), and
\end{subequations}
\begin{eqnarray}
\omega ^{2} &=&\frac{\left( T_{0}\Delta _{0}\right) ^{2}}{2T_{0}\Omega _{0}}=%
\frac{T_{0}\Delta _{0}^{2}}{2}\left( \left. \frac{d\Omega }{d\tau }\right|
_{\tau =0}\right) ^{-1}, \\
\varphi &=&\arg \Gamma \left( 1-i\frac{\omega ^{2}}{4}\right) -\arg \Gamma
\left( \frac{1}{2}-i\frac{\omega ^{2}}{4}\right) +\frac{\pi }{4}, \\
\psi &=&\eta _{d}(\tau _{p})-\frac{1}{2}\omega ^{2}\left( 1-\ln \frac{\omega
^{2}}{4}\right) +\frac{\pi }{4},  \notag \\
&&+\arg \Gamma \left( \frac{1}{2}-i\frac{\omega ^{2}}{4}\right) +\arg \Gamma
\left( 1-i\frac{\omega ^{2}}{4}\right) .
\end{eqnarray}%
Thus the lineshape $\left| B_{+}(\tau _{p})\right| ^{2}$ is
determined only by the dynamical phase and the first discontinuous
derivative of the pulse.

\begin{figure}[tb]
\centerline{\includegraphics[scale=0.75]{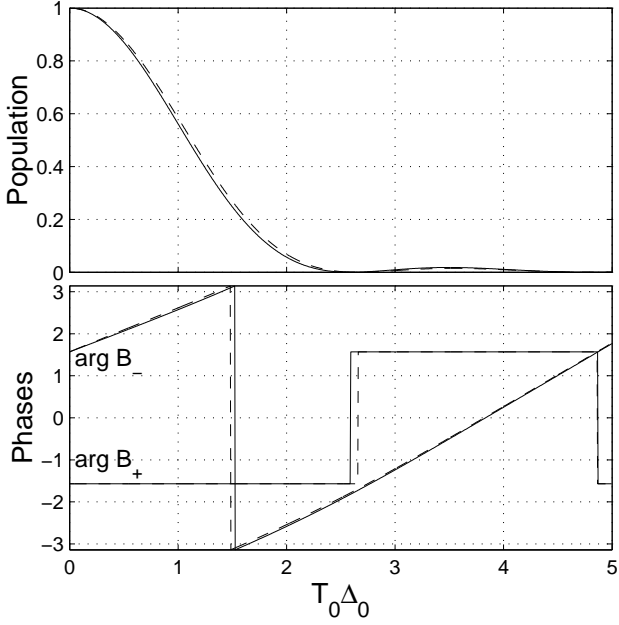}}
\caption{Lineshape $\left| B_{+}(\tau_p)\right| ^{2}$ [upper
frame; numerics (full lines) and using approximate formula
(\ref{finalBplus})], and phases $\arg A_{-}$ and $\arg A_{+}$ as
functions of $T_{0}\Delta_{0}$ for the trig-pulse (\ref{trig1})
and $T_{0}\Omega _{0}=\protect\pi /2$.} \label{Fig8}
\end{figure}

Figure 8 shows an example of the lineshape, accompanied with the phases, for
the trig-pulse
\begin{equation}
\Omega (\tau )=\Omega _{0}\sin \tau ,\quad 0\leq \tau \leq \pi ,
\label{trig1}
\end{equation}%
giving $\mathcal{A}=2T_{0}\Omega _{0}.$ We have chosen $T_{0}\Omega _{0}=\pi
/2$, i.e. a ''$\pi $-pulse'' which induces a complete population transfer
for $\Delta _{0}=0$ in this model. One can see that even for this $\pi $%
-pulse for which the condition of large area $\mathcal{A}\gg 1$ is valid
only very roughly, the numerical and analytical result (\ref{finalB_linear})
are very close. One can remark that as shown by Eq. (\ref{finalBplus}), the
phase of $B_{+}(\tau _{p})$ is $\pm \pi /2$ and does not depend on the
dynamical phase.

\subsection{Truncated pulses of power law endings}

We assume that the pulse starts and ends with power law rising and falling
around $\tau _{i}=0$ and $\tau _{f}=\tau _{p}\gg 1:$%
\begin{subequations}
\begin{eqnarray}
\Omega (\tau ) &\sim &\Omega _{0}\tau ^{n},\quad 1>\tau \geq 0, \\
\Omega (\tau ) &\sim &\Omega _{0}\left( \tau _{p}-\tau \right) ^{n},\quad
\tau _{p}-1<\tau \leq \tau _{p}
\end{eqnarray}%
with the discontinuous $n^{\text{th}}$ derivative
\end{subequations}
\begin{equation}
\Omega _{0}=\frac{1}{n!}\left. \frac{d^{n}\Omega }{d\tau ^{n}}\right| _{\tau
=0}.
\end{equation}
Using the generalization of the results for $n>1$ of the preceding Section,
we obtain
\begin{subequations}
\label{finalBn}
\begin{eqnarray}
B_{+}(\tau _{p}) &=&-2i\sqrt{p_{+}p_{-}}\sin \left( \chi _{-}+\chi _{+}+\eta
_{d}(\tau _{p})\right)  \label{finalBnplus} \\
B_{-}(\tau _{p}) &=&p_{-}e^{i\left( 2\chi _{-}+\eta _{d}(\tau _{p})\right)
}+p_{+}e^{-i\left( 2\chi _{+}+\eta _{d}(\tau _{p})\right) }
\end{eqnarray}%
with $p_{\pm }$, $\chi _{\pm }$ and $\eta _{d}$ respectively defined in Eqs.
(\ref{pplus_any}), (\ref{Chinmoins}), (\ref{Chinplus}) and (\ref{etad2}).
Thus the lineshape $P_{+}=\left| B_{+}(\tau _{p})\right| ^{2}$ is determined
only by the dynamical phase and the $n^{\text{th}}$ discontinuous derivative
of the pulse.

This allows to calculate approximately the lineshape for a trig-pulse
\end{subequations}
\begin{equation}
\Omega (\tau )=\Omega _{0}\sin ^{n}\tau ,\quad 0\leq \tau \leq \pi ,
\end{equation}%
where $\left. \frac{d^{n}\Omega }{d\tau ^{n}}\right| _{\tau =0}=\Omega
_{0}\tau ^{n}$.

\begin{figure}[tb]
\centerline{\includegraphics[scale=0.75]{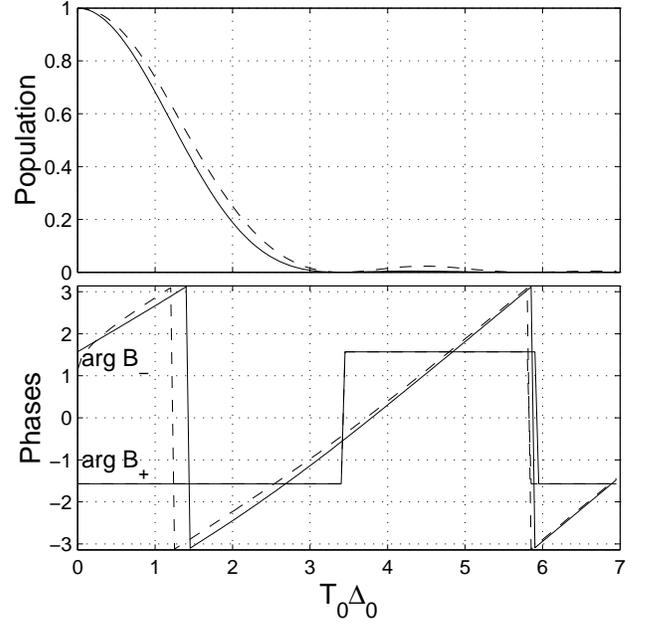}} \caption{Same
as Fig. 8 for the trig-pulse (\ref{trig2}) and $T_{0}\Omega
_{0}=2$ ($\protect\pi $-pulse).} \label{Fig9}
\end{figure}
\begin{figure}[tb]
\centerline{\includegraphics[scale=0.75]{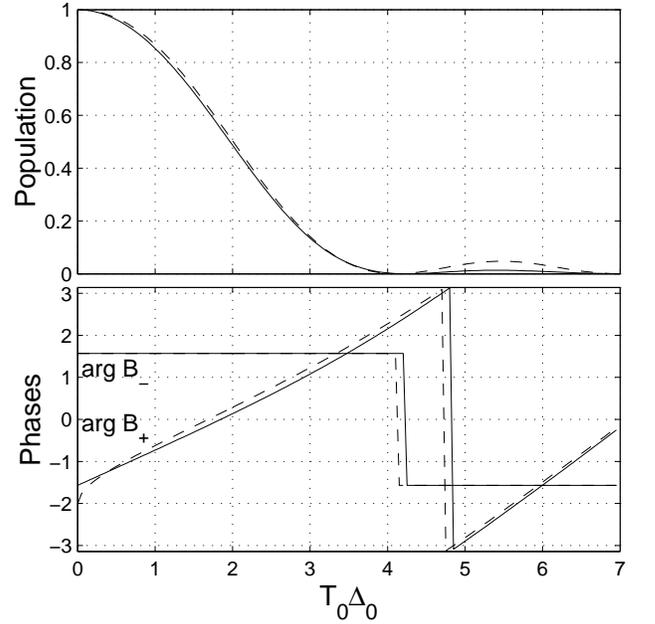}} \caption{Same
as Fig. 9 for $T_{0}\Omega _{0}=6$ ($3\protect\pi $-pulse).}
\label{Fig10}
\end{figure}

Figure 9 and 10 give two examples of the lineshape, accompanied with the
phases, for the trig-pulse
\begin{equation}
\Omega (\tau )=\Omega _{0}\sin ^{2}\tau ,\quad 0\leq \tau \leq \pi ,
\label{trig2}
\end{equation}%
giving $\mathcal{A}=T_{0}\Omega _{0}\pi /2$, respectively in ''$\pi $%
-pulse'' and ''3$\pi $-pulse'' conditions. The numerical and analytical
results are quite close.

\section{Application to Half-SCRAP}

It has been shown recently that two delayed pulsed lasers in an adiabatic
regime can be used to yield a coherent superposition of states. This process
has been named Half SCRAP \cite{HSCRAP}. It permits to create at the end of
the interaction a coherent superposition of states, whose amplitudes do not
depend on the dynamical phases and are thus robust with respect to the field
amplitudes. One uses two delayed lasers: a pump one-photon resonant laser
and an off-resonant Stark laser which allows to dynamically shift the
levels. The effective Hamiltonian is of the form (\ref{Ham1}) in the basis
of the dressed energies $\left| -;0\right\rangle $ (state $\left|
-\right\rangle $ dressed by 0 photon) and $\left| +;-1\right\rangle $ (state
$\left| +\right\rangle $ dressed by -1 photon), that are degenerate for the
exact one-photon resonance ($\Delta =0$) when $\Omega =0$.

We can interpret the process using the topology of the eigenenergy surfaces
as functions of the parameters $\Omega $ and $\Delta $, combined with a
local analysis of lifting (resp. creation) of quasi-degeneracy near the
start (resp. end) of the process.

The time dependence of the effective detuning $\Delta (t)=\Delta _{0}+S(t)$
is only due to Stark shifts which are induced by the laser pulses. The
process can be described by the diagram of the two surfaces
\begin{equation}
\lambda _{\pm }(\Omega ,\Delta )=\pm \frac{\hbar }{2}T_{0}\sqrt{\Omega
^{2}+\Delta ^{2}}  \label{surfs}
\end{equation}%
which represent the eigenenergies as functions of the instantaneous
effective Rabi frequency $\Omega $ and detuning $\Delta $ (see Fig. \ref%
{topo}). The associated eigenvectors can be written as
\begin{equation}
\left| \Phi _{-}\right\rangle =\left[
\begin{array}{c}
\cos \theta \\
-\sin \theta%
\end{array}%
\right] ,\quad \left| \Phi _{+}\right\rangle =\left[
\begin{array}{c}
\sin \theta \\
\cos \theta%
\end{array}%
\right]
\end{equation}%
with
\begin{equation}
\tan 2\theta =\frac{\Omega }{\Delta },\quad 0\leq \theta <\pi /2.
\end{equation}%
The surfaces display a conical intersection for $\Omega =0,\Delta =0$
induced by the crossing of the lines corresponding to states $\left|
-;0\right\rangle $ and $\left| +;-1\right\rangle $ for $\Omega =0$ and
various $\Delta $. We study a quasi-resonant process starting (and ending)
close to this conical intersection.

\begin{figure}[tb]
\centerline{\includegraphics[scale=0.75]{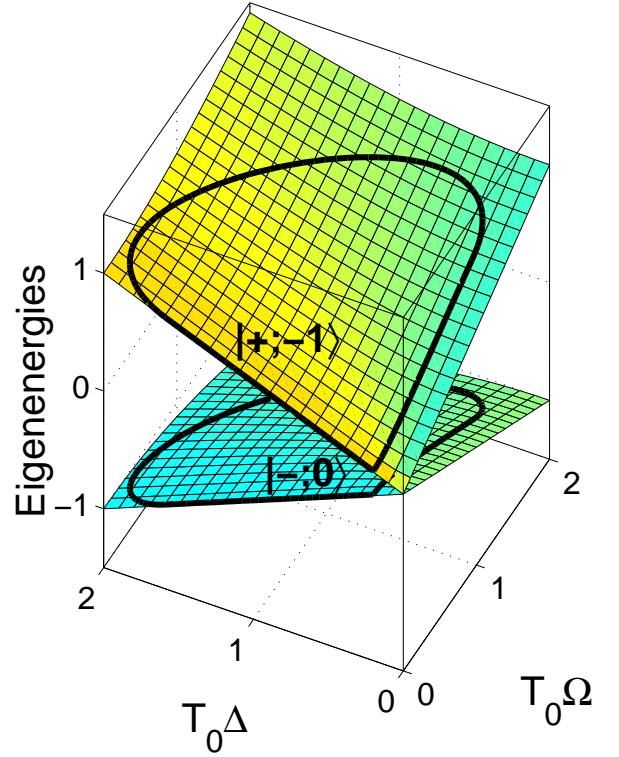}}
\caption{Surfaces of eigenenergies (\ref{surfs}) as functions of $%
T_{0}\Omega $ and $T_{0}\Delta $ (considered both positive). The lines in
the plane $\Omega=0$, connected to the respective lower and upper surfaces,
characterize the states respectively $\left| -;0\right\rangle $ and $\left|
+;-1\right\rangle $. One example of paths involving a lifting of
quasi-degeneracy is drawn.}
\label{topo}
\end{figure}
One essential point is that the result of the lifting (or creation) of
quasi-degeneracy depends on the direction of the lifting (or creation). We
can characterize two particular directions: In the $\Delta $ direction with $%
\Omega =0$, we have $\theta =0$, which means that the lifting and creation
of degeneracy occur trivially along a unique surface:
\begin{subequations}
\label{invliftDelta}
\begin{eqnarray}
\left| -;0\right\rangle &=&\left| \Phi _{-}\right\rangle ,  \label{liftDelta}
\\
\left| +;-1\right\rangle &=&\left| \Phi _{+}\right\rangle .
\end{eqnarray}%
In the $\Omega $ direction with $\Delta =0$, we have $\theta =\pi /4$, i.e
\end{subequations}
\begin{subequations}
\begin{eqnarray}
\left| -;0\right\rangle &=&\frac{1}{\sqrt{2}}\left( \left| \Phi
_{+}\right\rangle +\left| \Phi _{-}\right\rangle \right) ,  \label{liftOmega}
\\
\left| +;-1\right\rangle &=&\frac{1}{\sqrt{2}}\left( \left| \Phi
_{+}\right\rangle -\left| \Phi _{-}\right\rangle \right) ,
\end{eqnarray}%
which implies a lifting of degeneracy occuring along the lower and upper
surfaces with an equal sharing ($1/2$ if we consider the probabilities). In
other directions, the lifting of degeneracy results from a competition
between the parameters $\Delta $ and $\Omega $ and will give a mixing of the
two surfaces with non equal sharing in general.

The $\Omega $ direction with an arbitrary $\Delta $ will result in a lifting
of quasi-degeneracy as studied in the preceding sections.

The inverse process of creation of degeneracy also depends on the direction
in which the eigenstates involved in the dynamics arrive into the
degeneracy. In the $\Omega $ direction with $\Delta =0$, we have
\end{subequations}
\begin{subequations}
\begin{eqnarray}
\left| \Phi _{-}\right\rangle &=&\frac{1}{\sqrt{2}}\left( \left|
-;0\right\rangle -\left| +;-1\right\rangle \right)  \label{invliftOmega} \\
\left| \Phi _{+}\right\rangle &=&\frac{1}{\sqrt{2}}\left( \left|
-;0\right\rangle +\left| +;-1\right\rangle \right) ,
\end{eqnarray}

To obtain a coherent superposition of states, we have two possibilities:

(i) first lifting of degeneracy in the $\Delta $ direction with $\Omega =0$
[according to Eq. (\ref{liftDelta})] giving one single dressed state
involved in the dynamics; next adiabatic following on this dressed state
(along the lower surface) and finally creation of quasi-degeneracy in the $%
\Omega $ direction [according to Eq. (\ref{invliftOmega}) for the case of
exact resonance $\Delta =0$];

(ii) first lifting of quasi-degeneracy in the $\Omega $ direction [according
to Eq. (\ref{liftOmega}) for the case of exact resonance $\Delta =0$] giving
two dressed states involved in the dynamics; next \textit{independent}
adiabatic following on these two dressed states (along both the lower and
upper surfaces) and finally creation of degeneracy in the $\Delta $
direction with $\Omega =0$ [according to Eq (\ref{invliftDelta})].

These two cases are produced by two following different sequences of pulses:
respectively (i) first the Stark pulse and next the pump pulse (referred to
as Stark-pump sequence), (ii) first the pump pulse and next the Stark pulse
(referred to as pump-Stark sequence). (Both sequences require an overlapping
of the two pulses to induce adiabatic following.)

The dynamical phases coming from these two different sequences are not
identical. For the sequence Stark-pump at exact resonance $\Delta =0$, we
start with the lifting of degeneracy $\left| \phi \left( t_{i}\right)
\right\rangle =\left| \Phi _{-}\right\rangle $, followed by an adiabatic
passage
\end{subequations}
\begin{equation}
\left| \phi \left( t\right) \right\rangle =\text{e}^{-\text{i}%
\int_{t_{i}}^{t}ds\lambda _{-}(s)/\hbar }\left| \Phi _{-}\right\rangle ,
\end{equation}%
which leads at the final time $t_{f}$ to
\begin{equation}
\left| \phi \left( t_{f}\right) \right\rangle =\frac{1}{\sqrt{2}}\text{e}^{-%
\text{i}\int_{t_{i}}^{t_{f}}ds\lambda _{-}(s)/\hbar }\left( \left|
-\right\rangle -e^{-i\omega t}\left| +\right\rangle \right)  \label{S_pump}
\end{equation}%
(where a coherent state for the photon field has been considered.) For the
sequence pump-Stark at exact resonance $\Delta =0$, we start with the
lifting of degeneracy $\left| \phi \left( t_{i}\right) \right\rangle =\left(
\left| \Phi _{+}\right\rangle +\left| \Phi _{-}\right\rangle \right) /\sqrt{2%
}$. The dynamics is next characterized by an adiabatic passage along each
branch:
\begin{equation}
\left| \phi \left( t\right) \right\rangle =\frac{1}{\sqrt{2}}\left[ \text{e}%
^{-\text{i}\int_{t_{i}}^{t}ds\lambda _{+}(s)/\hbar }\left| \Phi
_{+}\right\rangle +\text{e}^{-\text{i}\int_{t_{i}}^{t}ds\lambda
_{-}(s)/\hbar }\left| \Phi _{-}\right\rangle \right] ,
\end{equation}%
which leads at time $t_{f}$ to
\begin{eqnarray}
\left| \phi \left( t_{f}\right) \right\rangle &=&\frac{1}{\sqrt{2}}\text{e}%
^{-\text{i}\int_{t_{i}}^{t_{f}}ds\lambda _{+}(s)/\hbar }  \notag \\
&&\times \left[ e^{-i\omega t}\left| +\right\rangle +\text{e}^{\text{i}%
\int_{t_{i}}^{t_{f}}\frac{ds}{\hbar }\left( \lambda _{+}(s)-\lambda
_{-}(s)\right) }\left| -\right\rangle \right] .  \label{pump_S}
\end{eqnarray}%
Thus the two sequences (with exact one-photon resonance) lead to the same
probabilities $1/2$ but with different phases. The pump-Stark sequence (\ref%
{pump_S}) leads to a coherent superposition of states with a (non-robust)
phase difference $\int_{t_{i}}^{t_{f}}ds\left( \lambda _{+}(s)-\lambda
_{-}(s)\right) /\hbar $, coinciding with the dynamical phase difference, in
addition to the optical phase.

\begin{figure}[tb]
\centerline{\includegraphics[scale=0.75]{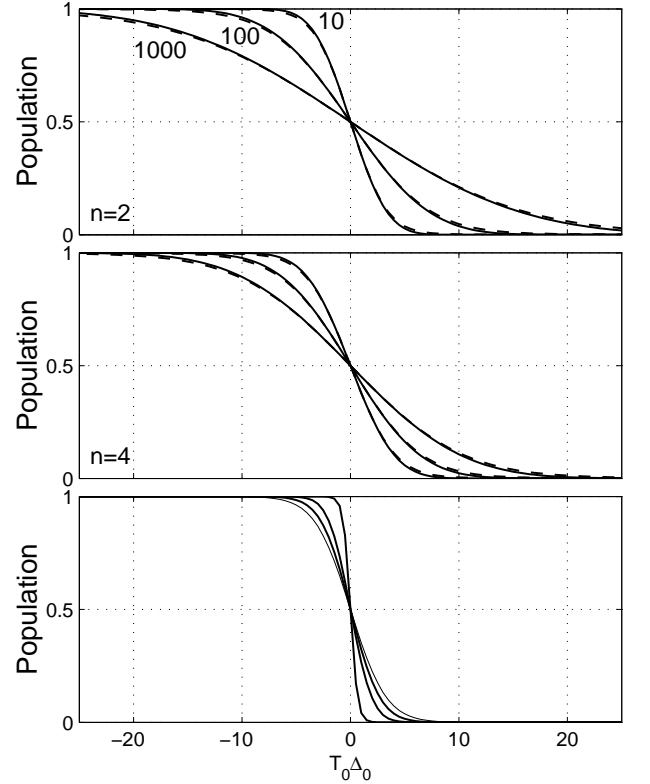}}
\caption{Half-SCRAP population transfers $P_{+}=\left|\langle +|\protect\phi%
(t_{f})\rangle\right| ^{2}$ for power rising of the pump laser (upper frame:
$n=2$, middle frame: $n=4$) with respectively $T_0\Omega_0=1000, 100, 10$
from (upper) left to right (indicated close to each curve) [numerics: full
lines, dashed lines: formulas (\ref{pplus_any})], and for Gaussian rising
with respectively $T_0\Omega_0=1000, 100, 10$ and exponential rising from
(upper) left to right (lower frame) as a function of $T_0\Delta_0$.}
\label{liftall}
\end{figure}

If one considers non exact one-photon resonance $\Delta \neq 0$, one
obtains, by lifting of quasi-degeneracy, additional phases and amplitudes
different from $1/\sqrt{2}$, that will depend on the shape of the pulses
according to the analysis of the preceding sections. Fig. \ref{liftall}
gathers populations of different coherent superposition of states given by
half-SCRAP for various pulse shapes. One can see that robustness with
respect to the detuning is better for power law rising than for the
exponential and Gaussian rising (better for small $n$ and large $\Omega _{0}$%
). Robustness with respect to the amplitude $\Omega _{0}$ is better for
smoother rising (and better for the exponential rising, which is independent
of $\Omega _{0}$, than for the Gaussian rising, which is weakly dependent on
$\Omega _{0}$).

\section{Conclusion}

In this article, we have analyzed the dynamics associated with lifting of
quasi-degeneracy with a constant detuning. This is the situation encountered
for one-photon quasi-resonance. The dynamics becomes quite different for $n-$%
photon quasi-resonance in effective two-level models, which leads to an
effective time-dependent detuning. Indeed, in this case the effective
Hamiltonian is, in the basis $\left\{ \left| -;0\right\rangle ,\left|
+;-n\right\rangle \right\} $, of the form%
\begin{equation}
\mathsf{H}(\tau )\simeq \frac{\hbar }{2}\left[
\begin{array}{cc}
-\alpha \mathcal{E}^{2}(\tau )-\Delta & \beta \mathcal{E}^{n}(\tau ) \\
\beta ^{\ast }\mathcal{E}^{n}(\tau ) & \alpha \mathcal{E}^{2}(\tau )+\Delta%
\end{array}%
\right]
\end{equation}%
with $\mathcal{E}(\tau )$ the field amplitude,$\ \alpha $ real (chosen
positive), $\beta =\left| \beta \right| e^{i\varphi }$, and where only the
leading second order (Stark shifts) have been kept in the diagonal. If $n=2$%
, all the terms of the matrix involving the field amplitude have the same
order which complicates the lifting of quasi-degeneracy. If $\Delta =0$, we
obtain the following lifting of degeneracy \cite{Guerin_PRA97,review}
\begin{subequations}
\begin{eqnarray}
\left| -;0\right\rangle &=&\sin \theta \left| \Phi _{+}\right\rangle +\cos
\theta \left| \Phi _{-}\right\rangle , \\
\left| +;-n\right\rangle &=&e^{i\varphi }\cos \theta \left| \Phi
_{+}\right\rangle -e^{i\varphi }\sin \theta \left| \Phi _{-}\right\rangle ,
\end{eqnarray}%
with
\end{subequations}
\begin{equation}
\tan 2\theta =\frac{\left| \beta \right| }{\alpha },\quad 0\leq \theta <\pi
/2.
\end{equation}%
If $n>2$, it is known that ``one can approximately compensate the Stark
shift''. This can be explained and extended as follows on the example for
the rising of the pulse: we separate the dynamics following the different
orders in $\mathcal{E}$ of the matrix: (i) the Stark shift at early times
shifts the diagonal elements (i.e. the dressed states in an effective way)
for small field amplitudes without transferring any population, and (ii) for
larger field amplitudes, the coupling lifts the resulting quasi-degeneracy.
This last step can be in principle analyzed by the tools presented in this
article. Thus, by adjusting $\Delta $, it is possible to cancel almost
completely or more generally partially the effect of the Stark shift,
leading to an arbitrary (in probability) superposition of states. The
complete cancellation is for example the key that permits to orient
molecules by adiabatic passage (in that case the Stark shift is of
exponential order) \cite{orient}.

The analysis of the lifting of quasi-degeneracy is more complicated in the
case $n=2$ since the Stark shifts and the lifting occur simultaneously,
leading to a lifting of quasi-degeneracy with a time-dependent detuning.
This requires the extension of the tools presented here.

The present analysis has allowed to recover the lineshape for secant
hyperbolic pulses (Rosen-Zener formula) and to determine quite precisely the
lineshape for trig-pulses. The lineshape for Gaussian pulses is to our
knowledge an open question.

\section{Acknowledgments}

We acknowledge support by INTAS 99-00019 and from Conseil R\'{e}gional de
Bourgogne. LY thanks l'Universit\'{e} de Bourgogne for several stays as
invited professor during which this work was accomplished.

\eject

\appendix

\section{Exact solution for a linearly rising coupling}

\label{linear}

The linearly rising (resp. falling) coupling problem can be solved exaclty
by expressing it as a Landau-Zener problem of finite duration, with a start
(resp. end) \textit{at} the avoided crossing, the so-called \textit{half
Landau-Zener problem} \cite{Vitanov}. In terms of time evolution operators,
we have to solve%
\begin{equation}
i\hbar \frac{\partial }{\partial \tau }U(\tau ,\tau _{0})=T_{0}\mathsf{H}%
(\tau )U(\tau ,\tau _{0}),\quad U(\tau _{0},\tau _{0})=\openone
\end{equation}%
and in the adiabatic basis%
\begin{equation}
i\hbar \frac{\partial }{\partial \tau }U_{A}(\tau ,\tau _{0})=\mathsf{H}%
_{A}(\tau )U_{A}(\tau ,\tau _{0}),\;U_{A}(\tau _{0},\tau _{0})=\openone
\end{equation}%
with
\begin{equation}
U_{A}(\tau ,\tau _{0})=\mathsf{R}^{\dagger }(\tau )U(\tau ,\tau _{0})\mathsf{%
R}(\tau _{0})
\end{equation}%
which can be written as%
\begin{equation}
U_{A}(\tau ,\tau _{0})=\left[
\begin{array}{cc}
U_{A}^{11}(\tau ,\tau _{0}) & U_{A}^{12}(\tau ,\tau _{0}) \\
-\left[ U_{A}^{12}(\tau ,\tau _{0})\right] ^{\ast } & \left[ U_{A}^{11}(\tau
,\tau _{0})\right] ^{\ast }%
\end{array}%
\right]  \label{UA}
\end{equation}%
[property also true for $U(\tau ,\tau _{0})$], since the Hamiltonian $%
\mathsf{H}(\tau )$ is of trace 0. ($\ast $ stands for the complex conjugate.)

We consider the linearly rising coupling

\begin{equation}
\Omega (\tau )=\left\{
\begin{array}{c}
\Omega _{0}\tau ,\;\tau \geqslant 0 \\
0,\;\tau <0%
\end{array}%
\right.  \label{app_linear+}
\end{equation}%
starting \ at $\tau =\tau _{i}=0$. We will denote $U_{A+}(\tau ,\tau _{0})$
[resp. $U_{A-}(\tau ,\tau _{0})$] the evolution operator asociated to the
rising (resp. falling) of the coupling

\subsection{Landau-Zener basis}

The amplitudes $\phi _{D}(\tau )$ of the states in the Landau-Zener basis
are obtained from the amplitudes of the bare states via the time-independent
unitary transformation $\mathsf{S}$:%
\begin{equation}
\phi _{LZ}(\tau )\equiv \left[
\begin{array}{c}
Z_{-}(\tau ) \\
Z_{+}(\tau )%
\end{array}%
\right] =\mathsf{S}^{\dagger }\phi (\tau ),
\end{equation}%
with
\begin{equation}
\mathsf{S}=\frac{1}{\sqrt{2}}\left[
\begin{array}{cc}
1 & 1 \\
-1 & 1%
\end{array}%
\right] ,  \label{LZtrans}
\end{equation}%
giving the Landau-Zener Hamiltonian
\begin{subequations}
\begin{align}
\mathsf{H}_{LZ}(\tau )& \equiv \mathsf{S}^{\dagger }\mathsf{H}(\tau )\mathsf{%
S} \\
& =\frac{\hbar }{2}\left[
\begin{array}{cc}
-\Omega (\tau ) & -\Delta _{0} \\
-\Delta _{0} & \Omega (\tau )%
\end{array}%
\right] .
\end{align}%
We introduce the dimensionless time variable
\end{subequations}
\begin{equation}
T(\tau )=\sqrt{\frac{T_{0}\Omega _{0}}{2}}\tau
\end{equation}%
that leads to the Schr\"{o}dinger equation
\begin{subequations}
\begin{align}
i\hbar \frac{\partial \tilde{\phi}_{LZ}}{\partial T}(T)& =\mathsf{\tilde{H}}%
_{LZ}(T)\tilde{\phi}_{LZ}(T), \\
\tilde{\phi}_{LZ}(T)& =\left[
\begin{array}{c}
\tilde{Z}_{-}(T) \\
\tilde{Z}_{+}(T)%
\end{array}%
\right] \equiv \phi _{LZ}(\tau )
\end{align}%
with the Hamiltonian
\end{subequations}
\begin{equation}
\mathsf{\tilde{H}}_{LZ}(T)=\hbar \left[
\begin{array}{cc}
-T & -\omega \\
-\omega & +T%
\end{array}%
\right] ,  \label{LZHam}
\end{equation}%
the dimensionless coupling
\begin{equation}
\omega =\frac{T_{0}\Delta _{0}}{\sqrt{2T_{0}\Omega _{0}}}
\end{equation}%
and the initial conditions at time%
\begin{equation}
T_{i}\equiv T(\tau _{i})=0.
\end{equation}

\subsection{Exact solution}

The problem is reduced to the finite Landau-Zener model with a start at $%
T=T_{i}=0$. Using the results of Ref. \cite{Vitanov} one can write the exact
solution of this problem:
\begin{equation}
\tilde{\phi}_{LZ}(T)=U_{LZ}\mathbf{(}T,T_{i})\tilde{\phi}_{LZ}(T_{i})
\end{equation}%
with
\begin{equation}
U_{LZ}(T,T_{i})=\mathsf{S}^{\dagger }U(\tau ,\tau _{i})\mathsf{S}
\end{equation}%
and the evolution operator
\begin{equation}
U_{LZ}\mathbf{(}T,T_{i})=\left[
\begin{array}{cc}
U_{LZ}^{11}\mathbf{(}T,T_{i}) & U_{LZ}^{12}\mathbf{(}T,T_{i}) \\
-\left[ U_{LZ}^{12}\mathbf{(}T,T_{i})\right] ^{\ast } & \left[ U_{LZ}^{11}%
\mathbf{(}T,T_{i})\right] ^{\ast }%
\end{array}%
\right] ,
\end{equation}%
whose matrix elements read
\begin{widetext}
\begin{subequations}
\begin{eqnarray}
U_{LZ}^{11}(T,T_{i}) &  = &\frac{\Gamma\left( 1-\frac{1}{2}i\omega
^{2}\right)  }{\sqrt{2\pi}}\left[ D_{i\omega^{2}/2}\left(  T\sqrt
{2}e^{-i\pi/4}\right) D_{-1+i\omega^{2}/2}\left(
T_{i}\sqrt{2}e^{i3\pi
/4}\right)  \right. \nonumber \\
& & \left.  +D_{i\omega^{2}/2}\left(  T \sqrt{2}e^{i3\pi/4}\right)
D_{-1+i\omega^{2}/2}\left(  T_{i}\sqrt{2}e^{-i\pi/4}\right)  \right]  ,\\
U_{LZ}^{12}(T,T_{i}) &  = &\frac{\Gamma\left(
1-\frac{1}{2}i\omega^{2}\right)  }{\omega\sqrt{\pi}}e^{i\pi/4}
\left[  D_{i\omega^{2}/2}\left( T \sqrt{2}e^{-i\pi/4}\right)
D_{i\omega^{2}/2}\left( T_{i}\sqrt {2}
e^{i3\pi/4}\right)  \right.  \nonumber\\
&  & \left.  -D_{i\omega^{2}/2}\left(  T
\sqrt{2}e^{i3\pi/4}\right)D_{i\omega^{2}/2} \left(
T_{i}\sqrt{2}e^{-i\pi/4}\right)  \right]  ,
\end{eqnarray}
\end{subequations}
where $D_{\nu}\left(  z\right)  $ represents the parabolic
cylinder function of order $\nu$ and argument $z$. Since in our
case $T_{i}=0$ and using
\begin{equation}
D_{\nu}\left(  0\right)  =\frac{2^{\nu/2}\sqrt{\pi}} {\Gamma\left(
\frac{1-\nu}{2}\right)  },
\end{equation}
we obtain
\begin{subequations}
\begin{eqnarray}
U_{LZ}^{11}(T,T_{i}) &  = &\frac{\Gamma\left(  1-\frac{1}{2}
i\omega^{2}\right)  2^{-1+i\omega^{2}/4}} {\Gamma\left(
1-\frac{1}{4}i\omega^{2}\right)  } \left[  D_{i\omega^{2}/2}\left(
T\sqrt{2}e^{-i\pi/4} \right)  +D_{i\omega^{2}/2}\left(
T\sqrt{2}e^{i3\pi/4}\right)
\right]  \\
U_{LZ}^{12}(T,T_{i}) &  = &\frac{2^{i\omega^{2}/4}e^{i\pi/4}
\Gamma\left(  1-\frac{1}{2}i\omega^{2}\right)  }
{\omega\Gamma\left(\frac{1}{2}-\frac{1}{4}i\omega^{2} \right)  }
\left[  D_{i\omega^{2}/2}\left(T\sqrt{2}e^{-i\pi/4}\right)
-D_{i\omega^{2}/2}\left(  T\sqrt{2}e^{i3\pi/4}\right)  \right]
\end{eqnarray}
\end{subequations}
\subsubsection{Asymptotics}
We calculate the asymptotic expansion for the evolution matrix
$U_{LZ}$ when the evolution is adiabatic, i.e. for $\tau\gtrsim1$
and for $T_0\Omega_0\gg1$ since we will consider intermediate or
small detunings. We will thus determine the asymptotic expansion
in the limit $T\gg 1$. Following Ref. \cite{Vitanov}, we have the
leading terms of the large-argument and large-order asymptotic
expansion
\begin{subequations}
\label{strong-coupling}
\begin{eqnarray}
D_{i\omega^{2}/2}\left(  T\sqrt{2}e^{-i\pi/4}\right)    &
\rightsquigarrow & \cos\vartheta(T) \exp\left[
\frac{\pi}{8}\omega^{2}
+i\eta\right]  \\
D_{i\omega^{2}/2}\left(  T\sqrt{2}e^{i3\pi/4}\right)    &
\rightsquigarrow & \cos\vartheta(T)  \exp\left[
-\frac{3\pi}{8}\omega^{2} +i\eta\right]
+\frac{\omega\sqrt{\pi}}{\Gamma\left(
1-\frac{1}{2}i\omega^{2}\right)  } \sin\vartheta(T)  \exp\left[
-\frac{\pi}{8} \omega^{2}-i\left(\eta+\frac{\pi}{4}\right) \right]
\end{eqnarray}
\end{subequations}
with
\begin{equation}
\tan2\vartheta(T)=\frac{\omega}{T} =\frac{\Delta_0}{\Omega_0\tau}
\end{equation}
i.e.
\begin{equation}
\vartheta(T)=\frac{\pi}{4}-\theta(\tau),
\end{equation}
and
\begin{equation}
\eta=-\frac{\omega^{2}}{4}+\frac{\omega^{2}}{2} \ln\left[
\frac{1}{\sqrt{2}}\left(  T +\sqrt{\omega^{2}+T^{2}}\right)
\right] +\frac{T}{2}\sqrt{\omega^{2}+T^{2}}.
\end{equation}
This asymptotic expansion (\ref{strong-coupling}) has been checked
numerically to be in fact valid when \emph{either} $\omega$ or $T$
is large. This will allow us to use it for any detuning
$\Delta_{0} $ in the adiabatic region, i.e. when $T\gg1$. Moreover
this asymptotic expansion is already a good approximation for $T
\sim3$. Using
\begin{equation}
\arg\left(  \frac{\Gamma\left(  1-\frac{1}{2}i\omega^{2}\right)  }
{\Gamma\left(  1-\frac{1}{4}i\omega^{2}\right)  } \right)
=\arg\Gamma\left(\frac{1}{2} -\frac{1}{4}i\omega^{2}\right)
-\frac{1}{2} \omega^{2} \ln2,
\end{equation}
\begin{equation}
 \left|  \Gamma\left(  1-i\alpha\right) \right|
=\sqrt{\frac{\pi\alpha} {\sinh\left( \pi\alpha\right)}},
\qquad\alpha\in \mathbb{R},
\end{equation}
and
\begin{equation}
\label{gamgam} \Gamma\left( z\right) \Gamma\left( z+1/2\right)
=\frac{\sqrt{\pi}}{2^{2z-1}} \Gamma(2z) , \qquad z\in \mathbb{C},
\end{equation}
we obtain
\begin{subequations}
\begin{eqnarray}
U_{LZ}^{11} \mathbf{(}T,T_{i})  &  \rightsquigarrow & a
e^{i\eta_{1}(T)} \cos \vartheta(T) +b
e^{-i\eta_{2}(T)}\sin\vartheta(T) \\
U_{LZ}^{12}\mathbf{(}T,T_{i}) & \rightsquigarrow & b
e^{i\eta_{2}(T)}
 \cos \vartheta(T) -ae^{-i\eta_{1}(T)}
\sin \vartheta(T)
\end{eqnarray}
with
\end{subequations}
\begin{subequations}
\begin{eqnarray}
\eta_{1}(T)  &  = &\widetilde\eta_{d}(T)
 -\frac{\omega^{2}}{4}\left(  1-\ln\frac{\omega ^{2}}{4}
\right)  +\arg\Gamma\left(  \frac{1}{2}-\frac{1}{4}
i\omega^{2}\right)  ,\\
\eta_{2}(T) &  = &\widetilde\eta_{d}(T)-\frac{\omega^{2}}{4}\left(
1-\ln \frac{\omega ^{2}}{4}\right) +\frac{\pi}{4}+\arg\Gamma\left(
1-\frac{1}{4}
i\omega^{2}\right)  ,\\
\widetilde\eta_{d}(T) &  \equiv &\int_{0}^{T}
\sqrt{\omega^{2}+T^{2}}dT=\frac{T}{2} \sqrt{\omega^{2}+T^{2}}
+\frac{\omega^{2}}{2}\ln\frac{T+\sqrt{\omega^{2}+T^{2}}}
{\omega},\\
& = & \eta_{d}(\tau)\equiv \frac{1}{\hbar}\int_0^\tau\lambda_+(\tau)d\tau\\
 a  &  = &\frac{1}{\sqrt{2}}
\sqrt{1+e^{-\pi\omega^{2}/2 }} ,\quad b =\frac{1}{\sqrt{2}}
\sqrt{1-e^{-\pi\omega^{2}/2}},
\end{eqnarray}

\end{subequations}
\end{widetext}where $\eta _{d}(T)$ corresponds to the dynamical phase
associated to the positive instantaneous eigenvalue $\sqrt{\omega ^{2}+T^{2}}
$ of the Landau-Zener Hamiltonian $\mathsf{\tilde{H}}_{LZ}(T)/\hbar $ of Eq.
(\ref{LZHam}).

\subsection{Time evolution operator for the adiabatic states}

The evolution operator in the basis of the adiabatic states can be written
as
\begin{equation}
U_{A+}(\tau ,0)=\mathsf{R}^{\dagger }\left( \tau \right) \mathsf{S}%
U_{LZ}(T(\tau ),0)\mathsf{S}^{\dagger }\mathsf{R}\left( 0\right) ,
\end{equation}%
with $\mathsf{R}\left( 0\right) =\openone$. Thus,
\begin{subequations}
\label{UAplus}
\begin{align}
U_{A+}^{11}(\tau ,0)& \rightsquigarrow \frac{1}{\sqrt{2}}\left[ ae^{i\eta
_{1}(T(\tau ))}+be^{i\eta _{2}(T(\tau ))}\right] , \\
U_{A+}^{12}(\tau ,0)& \rightsquigarrow \frac{1}{\sqrt{2}}\left[ -ae^{i\eta
_{1}(T(\tau ))}+be^{i\eta _{2}(T(\tau ))}\right] .
\end{align}

\section{Exact solution for exponentially rising coupling}

The case of exponentially rising and falling coupling can be solved
analytically in terms of the Kummer functions. Here we give the asymptotics
of the evolution operator in the adiabaticity region where the population of
the eigenstates is time independent.

We consider the rising coupling
\end{subequations}
\begin{equation}
\Omega (\tau )=\Omega _{0}e^{\tau }
\end{equation}%
with the initial condition at $\tau _{i}\rightarrow -\infty $.

\subsection{Evolution operator for the bare states}

It is convenient to introduce the new variable \cite{Delos,Delos2}
\begin{equation}
s(\tau )=T_{0}\int_{-\infty }^{\tau }\Omega (\tau ^{\prime })d\tau ^{\prime
}=T_{0}\Omega _{0}e^{\tau },
\end{equation}%
that corresponds to the partial dimensionless pulse area. In terms of this
new variable, the Schr\"{o}dinger equation reads

\begin{subequations}
\begin{align}
i\frac{\partial \tilde{\phi}}{\partial s}(s)& =\frac{1}{2}\left[
\begin{array}{cc}
-\Theta \left( s\right) & 1 \\
1 & \Theta \left( s\right)%
\end{array}%
\right] \tilde{\phi}(s), \\
\tilde{\phi}(s)& =\left[
\begin{array}{c}
\tilde{B}_{-}(s) \\
\tilde{B}_{+}(s)%
\end{array}%
\right] \equiv \phi (\tau )=\left[
\begin{array}{c}
B_{-}(\tau ) \\
B_{+}(\tau )%
\end{array}%
\right]
\end{align}%
where
\end{subequations}
\begin{equation}
\Theta \left( s\right) =\frac{\Delta \left[ \tau \left( s\right) \right] }{%
\Omega \left[ \tau \left( s\right) \right] }
\end{equation}%
is called the Stueckelberg variable. In our case $\Delta \left[ \tau \left(
s\right) \right] =\Delta _{0,}$ $\Omega \left[ \tau \left( s\right) \right]
=s/T_{0}$, i. e.
\begin{equation}
\Theta \left( s\right) =\frac{T_{0}\Delta _{0}}{s}.
\end{equation}%
The differential equation for $\tilde{B}_{-}(s)$ for an arbitrary function $%
\Theta \left( s\right) $ reads
\begin{equation}
\frac{d^{2}\tilde{B}_{-}}{ds^{2}}(s)=-\frac{1}{4}\left[ -2i\dot{\Theta}%
(s)+\Theta (s)^{2}+1\right] \tilde{B}_{-}(s),
\end{equation}%
which gives for the exponential coupling
\begin{equation}
\frac{d^{2}\tilde{B}_{-}}{ds^{2}}(s)=-\frac{1}{4}\left[ 2i\frac{T_{0}\Delta
_{0}}{s^{2}}+\frac{\left( T_{0}\Delta _{0}\right) ^{2}}{s^{2}}+1\right]
\tilde{B}_{-}(s).  \label{exp}
\end{equation}%
One starts from the time $\tau _{i}$ going to $-\infty .$ It means that $%
s_{i}=T_{0}\Omega _{0}e^{\tau _{i}}$ goes to zero (positive). The general
initial conditions for $\tilde{B}_{-}(s)$ are
\begin{subequations}
\begin{align}
\tilde{B}_{-}(s_{i})& =B_{-}(\tau _{i}), \\
\frac{d\tilde{B}_{-}}{ds}(s_{i})& =\frac{1}{2i}\left( B_{+}(\tau _{i})-\frac{%
T_{0}\Delta _{0}}{s}B_{-}(\tau _{i})\right) .
\end{align}%
We introduce the new function
\end{subequations}
\begin{equation}
C_{-}(s)=\tilde{B}_{-}(s)s^{-i\varpi /2}
\end{equation}%
with
\begin{equation}
\varpi =T_{0}\Delta _{0},
\end{equation}%
whose evolution equation is
\begin{equation}
\frac{d^{2}C_{-}}{ds^{2}}(s)+i\frac{\varpi }{s}\frac{dC_{-}}{ds}(s)+\frac{1}{%
4}C_{-}(s)=0.  \label{exprise}
\end{equation}%
The general initial conditions for $C_{-}(s)$ read
\begin{subequations}
\begin{align}
C_{-}(s_{i})& =B_{-}(\tau _{i})s_{i}^{-i\varpi /2}, \\
\frac{dC_{-}}{ds}(s_{i})& =\frac{1}{2i}B_{+}(\tau _{i})s_{i}^{-i\varpi /2}.
\end{align}%
Considering the particular initial condition $B_{-}(\tau _{i})=1,$ $%
B_{+}(\tau _{i})=0$ will give the components $U_{11}(\tau ,\tau
_{i})=B_{-}(\tau )$ and $U_{21}(\tau ,\tau _{i})=B_{+}(\tau )$ of the
evolution operator $U(\tau ,\tau _{i})$, which is enough to characterize
completely the evolution operator since we have $U_{22}(\tau ,\tau
_{i})=U_{11}^{\ast }(\tau ,\tau _{i})$ and $U_{12}(\tau ,\tau
_{i})=-U_{21}^{\ast }(\tau ,\tau _{i})$ (having the Hamiltonian $\mathsf{H}%
(\tau )$ of trace 0).

Eq. (\ref{exprise}) is a confluent hypergeometric equation, which for the
initial conditions $C_{-}(s_{i})=s_{i}^{-i\varpi /2},$ $\frac{dC_{-}(s_{i})}{%
ds}=0$ has the solution
\end{subequations}
\begin{equation}
C_{-}(s)=s_{i}^{-i\varpi /2}e^{-is/2}M(i\varpi /2,i\varpi ,is),
\end{equation}%
where $M(a,b,z)$ represents the Kummer function.

\subsection{Asymptotics}

We are interested in the limit $\Delta _{0}\ll \Omega _{0}$ with $%
T_{0}\Omega _{0}\gg 1$, which corresponds to the asymptotics for large $s\gg
\varpi $:%
\begin{equation}
M(i\varpi /2,i\varpi ,is)\rightsquigarrow \left( is\right) ^{-i\varpi /2}%
\frac{\Gamma \left( i\varpi \right) }{\Gamma \left( i\varpi /2\right) }%
\left( e^{is}+e^{-\pi \varpi /2}\right) .
\end{equation}%
Using Eq. (\ref{gamgam}) we obtain
\begin{equation}
C_{-}(s)\rightsquigarrow \left( s_{i}s\right) ^{-i\varpi /2}\frac{2^{i\varpi
}}{\sqrt{\pi }}\Gamma \left( \frac{1}{2}+i\frac{\varpi }{2}\right) \cosh
\left( \frac{\pi }{4}\varpi +i\frac{s}{2}\right) .
\end{equation}%
This leads to
\begin{subequations}
\begin{align}
U_{11}\mathbf{(}\tau ,\tau _{i})& =\left[ U_{22}\mathbf{(}\tau ,\tau _{i})%
\right] ^{\ast }=B_{-}(\tau )=\tilde{B}_{-}(s) \\
& \rightsquigarrow e^{i\xi }\frac{e^{is/2}+e^{-\left( \pi \varpi +is\right)
/2}}{\sqrt{2\left( 1+e^{-\pi \varpi }\right) }}, \\
U_{21}\mathbf{(}\tau ,\tau _{i})& =-\left[ U_{12}\mathbf{(}\tau ,\tau _{i})%
\right] ^{\ast }=B_{+}(\tau )=\tilde{B}_{+}(s) \\
& \rightsquigarrow e^{i\xi }\frac{e^{-\left( \pi \varpi +is\right)
/2}-e^{is/2}}{\sqrt{2\left( 1+e^{-\pi \varpi }\right) }},
\end{align}%
with
\end{subequations}
\begin{equation}
\xi =\arg \Gamma \left( \frac{1}{2}+i\frac{\varpi }{2}\right) +\varpi \ln 2-%
\frac{\varpi }{2}\ln s_{i},  \label{phase_ksi}
\end{equation}%
where we have used the definition of $\tilde{B}_{+}(s):\tilde{B}_{+}(s)=2i%
\frac{d\tilde{B}_{-}(s)}{ds}+\frac{\varpi }{s}\tilde{B}_{-}(s)$ and
\begin{equation}
\left| \Gamma \left( 1/2+i\alpha \right) \right| =\sqrt{\frac{\pi }{\cosh
\left( \pi \alpha \right) }},\qquad \alpha \in \mathbb{R}.
\end{equation}

\subsection{Evolution operator for the adiabatic states}

The evolution operator in the basis of the adiabatic states $U_{A}(\tau
,\tau _{i})=\mathsf{R}^{\dagger }(\tau )U(\tau ,\tau _{i})\mathsf{R}(\tau
_{i})$, $\mathsf{R}(\tau _{i})=\openone$, has to be considered for $\tau
\sim 1$ and $\Delta _{0}\ll \Omega _{0}$ which gives $\theta =\pi /4$ for
the transformation $\mathsf{R}^{\dagger }(\tau )$, according to Eq. (\ref%
{theta}). This leads to
\begin{subequations}
\label{UA_exp}
\begin{align}
U_{A+}^{11}(\tau ,\tau _{i})& =\frac{1}{\sqrt{2}}(U_{11}(\tau ,\tau
_{i})-U_{21}(\tau ,\tau _{i})) \\
& \rightsquigarrow \frac{e^{i\left( \xi +s/2\right) }}{\sqrt{1+e^{-\pi
\varpi }}}, \\
U_{A+}^{21}(\tau ,\tau _{i})& =\frac{1}{\sqrt{2}}(U_{11}(\tau ,\tau
_{i})+U_{21}(\tau ,\tau _{i})) \\
& \rightsquigarrow \frac{e^{-\pi \varpi /2}e^{i\left( \xi -s/2\right) }}{%
\sqrt{1+e^{-\pi \varpi }}}.
\end{align}%
If one starts with the initial condition $A_{-}(-\infty )=1,$ $A_{+}(-\infty
)=0,$ the amplitudes of the adiabatic states are $A_{-}(\tau
)=U_{A}^{11}(\tau ,\tau _{i})$ and $A_{+}(\tau )=U_{A}^{21}(\tau ,\tau _{i})$%
.

\section{Falling coupling and creation of degeneracy}

Using the time reversal symmetry, we calculate in this appendix, from the
assumed known evolution operator $U_{A+}(\tau ,\tau _{i})$ characterizing
the lifting of degeneracy starting at $\tau =\tau _{i}$, the evolution
operator $U_{A-}(\tau _{f},\tau )$ characterizing the creation of
degeneracy. We assume a process with a falling coupling starting in the
adiabatic region and ending at $\tau =\tau _{f}$. The coupling falling as a
power law (\ref{Rabi0_}) starts at $\tau \lesssim -1$ and ends at $\tau
=\tau _{f}=0$; the exponentially falling coupling (\ref{exp_falling}) starts
at $\tau \lesssim 0$ and ends at $\tau =\tau _{f}\rightarrow +\infty $. The
Hamiltonian associated to the pulse falling $\mathsf{H}_{-}(\tau )$ is
determined from the Hamiltonian associated to the pulse rising $\mathsf{H}%
_{+}(\tau )$ by $\mathsf{H}_{-}(\tau )=\mathsf{H}_{+}(-\tau )$. The relation
between the adiabatic Hamiltonians is
\end{subequations}
\begin{equation}
\mathsf{H}_{A-}(\tau )=\left[ \mathsf{H}_{A+}(-\tau )\right] ^{\ast }
\end{equation}%
with $\mathsf{H}_{A+}(\tau )$ and $\mathsf{H}_{A-}(\tau )$ the Hamiltonian
in the adiabatic states associated respectively to the rising and falling of
the coupling. The Schr\"{o}dinger equation reads in this case (for any $\tau
_{0}$)
\begin{equation}
i\hbar \frac{\partial }{\partial \tau }U_{A-}(\tau ,\tau _{0})=\mathsf{H}%
_{A-}(\tau )U_{A-}(\tau ,\tau _{0}).
\end{equation}%
After complex conjugation, using $\tau ^{\prime }=-\tau $, we obtain
\begin{equation}
U_{A-}(-\tau ^{\prime },\tau _{0})=\left[ U_{A+}(\tau ^{\prime },\tau
_{0}^{\prime })\right] ^{\ast }
\end{equation}%
with $\tau _{0}^{\prime }=-\tau _{0}$ in order to satisfy $U_{A-}(\tau
_{0},\tau _{0})=\left[ U_{A+}(-\tau _{0},\tau _{0}^{\prime })\right] ^{\ast
}=\openone.$ Thus
\begin{equation}
U_{A-}(\tau _{0},-\tau ^{\prime })=\left[ U_{A-}(-\tau ^{\prime },\tau _{0})%
\right] ^{\dagger }=\left[ U_{A+}(\tau ^{\prime },-\tau _{0})\right] ^{t},
\end{equation}%
where $t$ stands for the transposed. Choosing $\tau _{0}=\tau _{f}$, we get
for any $\tau $ in the adiabatic region
\begin{subequations}
\label{UAmoins}
\begin{align}
U_{A-}(\tau _{f},\tau )& =\left[ U_{A+}(-\tau ,-\tau _{f})\right] ^{t} \\
& =\left[
\begin{array}{cc}
U_{A+}^{11}(-\tau ,-\tau _{f}) & -\left[ U_{A+}^{12}(-\tau ,-\tau _{f})%
\right] ^{\ast } \\
U_{A+}^{12}(-\tau ,-\tau _{f}) & \left[ U_{A+}^{11}(-\tau ,-\tau _{f})\right]
^{\ast }%
\end{array}%
\right]
\end{align}%
with $\tau _{f}=0$ for the coupling falling as a power law (\ref{Rabi0_})
and $\tau _{f}\rightarrow +\infty $ for the exponentially falling coupling (%
\ref{exp_falling}).

If the coupling falling as a power law ends at $\tau _{f}\neq 0:$%
\end{subequations}
\begin{equation}
\Omega (\tau )=\left\{
\begin{array}{c}
-\Omega _{0}\left( \tau -\tau _{f}\right) ^{n},\;\tau \leq \tau _{f} \\
0,\;\tau >\tau _{f}%
\end{array}%
\right. ,
\end{equation}%
we obtain using the same method%
\begin{equation}
U_{A-}(\tau _{f},\tau )=\left[ U_{A+}(-\tau +2\tau _{f},\tau _{f})\right]
^{t}.  \label{UAmoins_}
\end{equation}

\section{Asymptotics for large detuning: complements}

In this appendix, we show formula (\ref{In}), starting from the integral (%
\ref{Inint})
\begin{equation}
I_{n}\equiv \left( \frac{-1}{i\alpha _{n}}\right) ^{n}\int_{0}^{\infty
}\partial _{\omega }^{n}g_{n}(\omega )e^{i\alpha _{n}\omega }.
\end{equation}%
We first notice that $\partial _{\omega }^{n}g_{n}(\omega )$ are odd
functions for any $n$. Thus
\begin{subequations}
\label{InIpn}
\begin{align}
\text{Im}I_{n}& =\frac{1}{2}\text{Im}I_{n}^{\prime }\quad \text{for }n\text{
even,} \\
\text{Re}I_{n}& =\frac{1}{2}\text{Re}I_{n}^{\prime }\quad \text{for }n\text{
odd,}
\end{align}%
with
\end{subequations}
\begin{equation}
I_{n}^{\prime }=\left( \frac{-1}{i\alpha _{n}}\right) ^{n}\int_{-\infty
}^{+\infty }\partial _{\omega }^{n}g_{n}(\omega )e^{i\alpha _{n}\omega }.
\end{equation}%
Thus standard techniques of contour integration allow to calculate with high
accuracy $\text{Im}I_{n}$ for $n$ even and $\text{Re}I_{n}$ for $n$ odd as
follows. We have to evaluate
\begin{equation}
I_{n}^{\prime }=\left( \frac{-1}{i\alpha _{n}}\right) ^{n}\oint \partial
_{\omega }^{n}g_{n}(\omega )e^{i\alpha _{n}\omega }
\end{equation}%
In the upper half complex plane, $g_{n}\left[ \omega (x)\right] $ has $n$
poles (with $x$ extended in the complex plane)
\begin{equation}
x_{c}^{(k)}=e^{i\frac{\pi }{n}\left( \frac{1}{2}+k\right) },\quad
k=0,1,\cdots ,n-1,
\end{equation}%
that are associated to poles
\begin{subequations}
\begin{align}
\omega _{c}^{(k)}& \equiv \int_{0}^{x_{c}^{(k)}}\sqrt{1+u^{2n}}du \\
& =b_{n}x_{c}^{(k)}
\end{align}%
of order one in the variable $\omega $ with
\end{subequations}
\begin{equation}
b_{n}\equiv \int_{0}^{1}\sqrt{1-x^{2n}}dx=\frac{1}{4n}\sqrt{\pi }\frac{%
\Gamma \left( \frac{1}{2n}\right) }{\Gamma \left( \frac{3n+1}{2n}\right) }.
\end{equation}%
We evaluate the function $g_{n}(\omega )$ around each of these poles as
\begin{equation}
g_{n}^{(k)}(\omega )\approx -\frac{1}{6}\frac{x_{c}^{(k)}}{\omega -\omega
_{c}^{(k)}},
\end{equation}%
using the relations around the poles
\begin{equation}
1+x^{2n}\approx \frac{-2n}{x_{c}^{(k)}}\left( x-x_{c}^{(k)}\right)
\end{equation}%
and
\begin{equation}
\omega \approx \omega _{c}^{(k)}+\frac{2}{3}\sqrt{\frac{-2n}{x_{c}^{(k)}}}%
\left( x-x_{c}^{(k)}\right) ^{3/2}.
\end{equation}%
This simply leads to
\begin{equation}
\partial _{\omega }^{n}g_{n}^{(k)}(\omega )\approx \left( -1\right) ^{n+1}%
\frac{n!}{6}\frac{\left( x_{c}^{(k)}\right) ^{n}}{\left( \omega -\omega
_{c}^{(k)}\right) ^{n+1}}.  \label{aproxdg}
\end{equation}%
Hence, $\partial _{\omega }^{n}g_{n}^{(k)}(\omega )$ having $n$ poles of
order $n+1$, we obtain
\begin{equation}
I_{n}^{\prime }=2\pi i\sum_{\text{Im}(\omega )>0}\text{Res}\left[ \partial
_{\omega }^{n}g_{n}(\omega )e^{i\alpha _{n}\omega }\right]
\end{equation}%
with the residue for each pole
\begin{widetext}
\begin{subequations}
\begin{eqnarray}
\text{Res}\left[
\partial_{\omega}^{n}
g_{n}(\omega)e^{i\alpha_{n}\omega }\right] & \equiv &\frac{1}{n!}
\times\lim_{\omega\rightarrow\omega_{c}^{(k)}}\partial_{\omega}^{n}
\left[ \left(  \omega-\omega_{c}^{(k)}\right)  ^{n+1}
\partial_{\omega}^{n}
g_{n}(\omega)e^{i\alpha_{n}\omega}
\right]  \\
& = &\frac{1} {6}\left( -1\right) ^{n+1} \left( i\alpha_{n}
x_{c}^{(k)}\right) ^{n} e^{i\alpha_{n}b_{n} x_{c}^{(k)}},
\end{eqnarray}

\end{subequations}
\end{widetext}giving
\begin{equation}
I_{n}^{\prime }\approx \frac{\pi }{3}\sum_{k=0}^{n-1}\left( -1\right)
^{k}e^{i\alpha _{n}b_{n}x_{c}^{(k)}}.
\end{equation}%
This leads to
\begin{equation}
I_{n}^{\prime }\approx \left\{
\begin{array}{c}
\frac{2\pi }{3}\sum_{k=0}^{n/2-1}i\text{Im}\left( F_{k}\right) \text{,\quad
for even }n \\
\frac{2\pi }{3}\left[ \frac{1}{2}(-1)^{\frac{n-1}{2}}e^{-\alpha
_{n}b_{n}}+\sum_{k=0}^{\frac{n-3}{2}}\text{Re}\left( F_{k}\right) \right]
\text{,\ for odd }n%
\end{array}%
\right.
\end{equation}%
with
\begin{equation}
F_{k}=(-1)^{k}e^{-\alpha _{n}b_{n}\sin \left[ \frac{\pi }{n}\left( k+\frac{1%
}{2}\right) \right] }e^{i\alpha _{n}b_{n}\cos \left[ \frac{\pi }{n}\left( k+%
\frac{1}{2}\right) \right] }.
\end{equation}%
We can remark that this result is approximative due to the approximation (%
\ref{aproxdg}). It is more accurate for larger $\alpha _{n}$. This allows to
calculate $\text{Im}I_{n}$ for $n$ even and $\text{Re}I_{n}$ for $n$ odd as
prescribed by (\ref{InIpn}). We have additionally found numerically that
extending $I_{n}^{\prime }$ as follows
\begin{equation}
I_{n}^{\prime \prime }\equiv \left\{
\begin{array}{c}
\frac{2\pi }{3}\sum_{k=0}^{n/2-1}F_{k}\text{,\quad for even }n \\
\frac{2\pi }{3}\left[ \frac{1}{2}(-1)^{\frac{n-1}{2}}e^{-\alpha
_{n}b_{n}}+\sum_{k=0}^{\frac{n-3}{2}}F_{k}\right] \text{,\ for odd }n%
\end{array}%
\right.
\end{equation}%
and setting for all $n$
\begin{equation}
I_{n}\approx \frac{1}{2}I_{n}^{\prime \prime },
\end{equation}%
which leads to Eq. (\ref{In}), allows to calculate $I_{n}$ with a good
approximation for sufficiently large values of $\alpha _{n}$.


\begin{references}
\bibitem{entang}
D. Bouwmeester, A. Ekert, and A. Zeilinger, {\it The Physics of
Quantum Information: Quantum Criptography, Quantum Teleportation,
Quantum Computation} (Springer Verlag, Berlin, 2000).

\bibitem{Holthaus}
M. Holthaus and B. Just, Phys. Rev. A {\bf 49}, 1950  (1994).

\bibitem{Guerin_PRA97}
S. {Gu\'{e}rin} and H. R. Jauslin, Phys. Rev. A {\bf 55}, 1262
(1997).

\bibitem{review}
S. {Gu\'{e}rin} and H. R. Jauslin, Adv. Chem. Phys. {\bf 125}, 147
(2003).

\bibitem{Just}
B. Just, J. Manz, G. K. Paramonov Chem. Phys. Lett. {\bf 193}, 429
(1992).

\bibitem{Korolkov}
M. V. Korolkov, J. Manz, G. K. Paramonov Chem. Phys. {\bf 217},
341 (1997).

\bibitem{Berman}
P. R. Berman, L. Yan, K.-H. Chiam, and R. Sung Phys. Rev. A {\bf
57}, 79 (1998) and references therein.

\bibitem{Guerin_PR00}
S. Gu\'{e}rin, L. P. Yatsenko and H. R. Jauslin, Phys. Rev. A {\bf
63}, R031403 (2001).

\bibitem{Yatsenko_PR01}
L. P. Yatsenko, S. Gu\'{e}rin and H. R. Jauslin, Phys. Rev. A {\bf
65}, 043407 (2002).

\bibitem{Yatsenko_PR99}
L. P. Yatsenko, B. W. Shore, T. Halfmann, K. Bergmann, and A.
Vardi, Phys. Rev. A {\bf 60}, R4237 (1999).

\bibitem{Rickes}
T. Rickes, L. P. Yatsenko, S. Steuerwald, T. Halfmann, B. W.
Shore, N. V. Vitanov and K. Bergmann, J. Chem. Phys. {\bf 113},
534 (2000).

\bibitem{Landau}
L. D. Landau, Phys. Z. Sowjetunion {\bf 2}, 46 (1932).

\bibitem{Zener}
C. Zener, Proc. R. Soc. London A\ {\bf 137}, 696 (1932).

\bibitem{Dykhne}
A. M. Dykhne, Sov. Phys. JETP {\bf 14}, 941 (1962).

\bibitem{Davis}
J. P. Davis and P. Pechukas, J. Chem. Phys. {\bf 64}, 3129 (1976).

\bibitem{Vitanov}
N. V. Vitanov and B.M. Garraway, Phys. Rev. A {\bf 53}, 4288,
(1996).

\bibitem{Allen}
L. Allen and J. H. Eberly, {\em Optical Resonance and Two-Level
Atoms} (Dover, New York, 1987).

\bibitem{Shore}
B.~W. Shore, {\em The Theory of Coherent Atomic Excitation}
(Wiley, New York, 1990).

\bibitem{Garrido}
L. M. Garrido and F. J. Sancho, Physica 28, 553 (1962).

\bibitem{Sancho}
F. J. Sancho, Proc. Phys. Soc. 89, 1 (1966).

\bibitem{RosenZener}
N. Rosen and C. Zener, Phys. Rev. {\bf 40}, 502 (1932).

\bibitem{HSCRAP}
L. P. Yatsenko, N. V. Vitanov, B. W. Shore, T. Rickes, and K.
Bergmann, Opt. Commun. {\bf 204}, 413 (2002).

\bibitem{Delos}
J. B. Delos and W. R. Thorson, Phys. Rev. A {\bf 6}, 728 (1972).

\bibitem{Delos2}
T. R. Dinterman and J. B.Delos, Phys. Rev. A {\bf 15}, 463 (1977).

\bibitem{orient}
S. Gu\'erin, L. P. Yatsenko, H. R. Jauslin, O. Faucher and B.
Lavorel, Phys. Rev. Lett. {\bf 88}, 233601  (2002).

\end{references}
\end{document}